\documentclass[sn-mathphys-num]{sn-jnl}

\usepackage{graphicx}%
\usepackage{multirow}%
\usepackage{amsmath,amssymb,amsfonts}%
\usepackage{amsthm}%
\usepackage{mathrsfs}%
\usepackage[title]{appendix}%
\usepackage{xcolor}%
\usepackage{textcomp}%
\usepackage{manyfoot}%
\usepackage{booktabs}%
\usepackage{algorithm}%
\usepackage{algorithmicx}%
\usepackage{algpseudocode}%
\usepackage{listings}%
\usepackage[matrix,arrow]{xy}%

\theoremstyle{thmstyleone}%
\newtheorem*{theorem*}{Theorem}
\newtheorem{theorem}{Theorem}[section]
\newtheorem{proposition}{Proposition}[section]%
\newtheorem{lemma}{Lemma}[section]%
\newtheorem{corollary}{Corollary}[section]%

\theoremstyle{thmstyletwo}%
\newtheorem{example}{Example}[section]%
\newtheorem{remark}{Remark}[section]%

\theoremstyle{thmstylethree}%
\newtheorem{definition}{Definition}[section]%

\begin{document}

\title[Article title]{Higher Courant-Dorfman algebras and associated higher Poisson vertex algebras}


\author[1]{\fnm{Ryo} \sur{Hayami}}\email{ryo-hayami@nagano.ac.jp}

\affil[1]{\orgdiv{} \orgname{Nagano University}, \orgaddress{\city{Ueda}, \postcode{386-1298}, \country{Japan}}}


\abstract{In this paper, we consider a higher notion of the relation between Courant-Dorfman algebras and Poisson vertex algebras in terms of dg symplectic geometry. We define a higher notion of Courant-Dorfman algebras, and study the relationship with graded symplectic geometry. In particular, we give graded Poisson algebras of degree $-n$ in the non-degenerate case. For higher Courant-Dorfman algebras coming from finite-dimensional vector bundles, they coincide with the algebras of functions of the associated dg symplectic manifolds of degree $n$.  We define a higher notion of Lie conformal algebras and Poisson vertex algebras, and give a higher (weak) Courant-Dorfman algebraic structure arising from them. Moreover, we prove that the higher Lie conformal algebras and higher Poisson vertex algebras have properties like Lie conformal algebras and Poisson vertex algebras. As an example, we get an algebraic description of BFV current algebras. This description can be seen as a higher analog of the relation between Poisson vertex algebras and Alekseev-Strobl current algebras.}




\maketitle

\section{Introduction}\label{sec1}

A Courant algebroid is a 4-tuple ($E,\rho,\langle,\rangle,[,]$) where $E$ is a vector bundle over a smooth manifold $M$, $\rho$ is an anchor map to the tangent bundle, $\langle,\rangle$ is a non-degenerate metric, and $[,]$ is a Courant bracket on the sections of the bundle, satisfying a set of compatibility conditions. It first appeared in \cite{C90} as the generalized tangent bundle $TM\oplus T^{*}M$ with the natural projection $\rho:TM\oplus T^{*}M\rightarrow TM$, the natural pairing $\langle.\rangle$ and the Dorfman bracket $[,]$. The general definition was given in \cite{liu1997manin} to generalize the double of Lie bialgebroids (Lie algebroid analog of Lie bialgebras\cite{vg1990hamiltonian}). We can get a map $d:C^{\infty}(M)\rightarrow\Gamma(E)$ by defining $\langle df,e\rangle=\rho(e)f$ for $f\in C^{\infty}(M),e\in\Gamma(E)$. Courant algebroids play important roles in some areas of mathematics and physics, for example, generalized geometries\cite{Gualtieri:2003dx}, T-dualities\cite{cavalcanti2011generalized}, topological sigma models\cite{roytenberg2007aksz},supergravity\cite{coimbra2011supergravity}, and double field theories\cite{vaisman2013towards}. Moreover, there is a one-to-one correspondence between the isomorphism classes of differential-graded (dg for short) symplectic manifolds of degree 2 and isomorphism classes of Courant algebroids\cite{roytenberg2002structure}.

A Courant-Dorfman algebra, whose definition was given in \cite{roytenberg2009courant}, is a 5-tuple ($R,E,\partial,\langle,\rangle,[,]$), where $R$ is a commutative algebra, $E$ is an $R$-module, $\langle ,\rangle:E\otimes E\rightarrow R$ is a symmetric bilinear form, $\partial:R\rightarrow E$ is a derivation, and $[,]:E\otimes E\rightarrow E$ is a Dorfman bracket, satisfying a set of compatibility conditions. A Courant algebroid gives a Courant-Dorfman algebra via ($C^{\infty}(M),\Gamma(E),d,\langle,\rangle,[,]$). Courant-Dorfman algebras generalize Courant algebroids in two directions: first allowing for more general commutative algebras $R$ and modules $E$ than algebras of smooth functions and modules of smooth sections, and second allowing for degenerate $\langle,\rangle$.  The relation between Courant-Dorfman algebras and Poisson vertex algebras was found in the context of current algebras\cite{ekstrand2011courant}\cite{ekstrand2011going}. 

Current algebras are Poisson algebras consisting of functions on mapping spaces. In classical field theories, a Poisson algebraic structure of currents plays important roles when we consider symmetries of fields. The most basic example is a Kac-Moody algebra, which is the Lie algebraic structure on $\mathrm{Map}(S^{1},G)$, where $G$ is a Lie group. Let $\mathfrak{g}$ be the Lie algebra of $G$ and $e_{a}$ be generators of $\mathfrak{g}$ such that $[e_{a},e_{b}]=f^{c}_{ab}e_{c}$. The bracket is of the form
\begin{equation}
\label{KM}
\{J_{a}(\sigma),J_{b}(\sigma')\}=f^{c}_{ab}J_{c}(\sigma)\delta(\sigma-\sigma')+k\delta_{ab}\delta'(\sigma-\sigma'),
\end{equation}
where $k$ is a constant. The algebra plays important roles as the symmetry of the Wess-Zumino-Witten model, 2-dimensional conformal invariant sigma model whose target space is a Lie group\cite{knizhnik1984current}.

Alekseev and Strobl observed that there was more general current algebra whose source manifold was $S^{1}$ but target manifold was a general smooth manifold\cite{alekseev2005current}. Let $M$ be a smooth manifold and choose a vector field $v=v^{i}(x)\partial_{i}$ and a 1-form $\alpha=\alpha_{i}(x)dx^{i}$ on $M$. We associate to them a current,
\begin{equation}
J_{(v,\alpha)}(\sigma)=v^{i}(x(\sigma))p_{i}(\sigma)+\alpha_{i}(x(\sigma))\partial_{\sigma}x^{i}(\sigma).
\end{equation}

 The Poisson bracket of these currents is of the form,
\begin{equation}
\{J_{(v,\alpha)}(\sigma),J_{(u,\beta)}(\sigma')\}=J_{[(v,\alpha),(u,\beta)]}(\sigma)\delta(\sigma-\sigma')+\langle(v,\alpha),(u,\beta)\rangle(\sigma)\delta'(\sigma-\sigma'),
\end{equation}
where $u,v$ is a vector field on $M$, $\alpha,\beta$ is a 1-form on $M$, $[(v,\alpha),(u,\beta)]=([v,u],L_{v}\beta-\iota_{u}d\alpha)$ is the Dorfman bracket on the generalized tangent bundle $TM\oplus T^{*}M$ and $\langle(v,\alpha),(u,\beta)\rangle=\iota_{u}\alpha+\iota_{v}\beta$. Let $M$ be a Lie group, and consider an Alekseev-Strobl current of the form
\begin{equation}
J_{(e,-\frac{k}{4\pi}g^{-1}dg)}=p(\sigma)-\frac{k}{4\pi}g^{-1}(\sigma)\partial_{\sigma}g(\sigma).
\end{equation}
The Poisson bracket of these currents is Kac-Moody algebra. Alekseev-Strobl currents appear in the description of symmetries of 2-dimensional $\sigma$-models. 

Inspired by \cite{alekseev2005current}, Ekstrand and Zabzine studied the algebraic structure underlying more general current algebras on loop spaces,\cite{ekstrand2011courant} They found that a weak notion of Courant-Dorfman algebras (weak Courant-Dorfman algebras) appears when we consider the Poisson bracket of currents. In \cite{ekstrand2011going}, (weak) Courant-Dorfman algebras were derived using the language of Lie conformal algebras (LCA for short) and Poisson vertex algebras (PVA for short).

A Lie conformal algebra is a module with a $\lambda$-bracket satisfying some conditions like a Lie algebra, and a Poisson vertex algebra is defined as an algebra which has a structure of a Lie conformal algebra and satisfies the Leibniz rule. The relation with the Poisson bracket of currents was investigated in \cite{barakat2009poisson}. We can get a Lie conformal algebra from the Poisson bracket of currents, and we can get a Poisson vertex algebra by taking into account the multiplication of currents. Poisson vertex algebras can be seen as an algebraic generalization of the Poisson brackets on loop spaces, while Lie conformal algebras can be seen as an algebraic generalization of the Lie brackets on loop spaces. In \cite{ekstrand2011going}, Ekstrand derived weak Courant-Dorfman algebras from Lie conformal algebras and showed that the graded Poisson vertex algebras generated by elements of degree 0 and 1 are in one-to one correspondence with the Courant-Dorfman algebras. 

The above discussions are summarized as follows.

\begin{equation}
  \xymatrix{
    \fbox{degree 2 dg symplectic manifolds} \ar@{<->}[r]^-{1-to-1}  & \fbox{Courant algebroids}
  }
\end{equation}

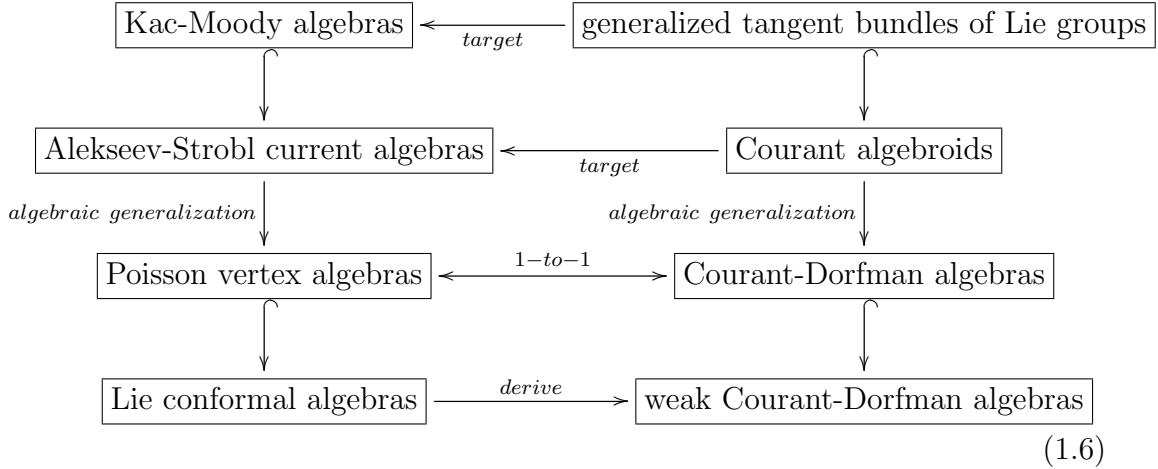
\begin{equation}
\label{2}
  \xymatrix{
    \fbox{Kac-Moody algebras}\ar@{^{(}->}[d] & \fbox{generalized tangent bundles of Lie groups}  \ar[l]^-{target} \ar@{^{(}->}[d] \\
    \fbox{Alekseev-Strobl current algebras} \ar@{^{(}->}[d] & \fbox{Courant algebroids} \ar[l]^-{target} \ar@{^{(}->}[d] \\
    \fbox{Poisson vertex algebras} \ar@{<->}[r]^-{1-to-1}  \ar@{^{(}->}[d] & \fbox{Courant-Dorfman algebras} \ar@{^{(}->}[d]\\
    \fbox{Lie conformal algebras}  \ar[r]^-{derive}& \fbox{weak Courant-Dorfman algebras} 
  }
\end{equation}
In this diagram, ``target'' means that the right object is the structure of the target of the left object. 

Courant algebroids are in one-to-one correspondence with degree 2 dg symplectic manifolds, and Alekseev-Strobl current algebras can be described in the language of dg symplectic geometry\cite{ikeda2013current}. Moreover, Poisson algebras on the mapping space whose source manifold is higher dimensional were constructed.(For example, \cite{bonelli2005current},\cite{hatsuda2012canonical}) and a general framework explaining these current algebras were given using dg symplectic geometry.\cite{ikeda2013currentdg},\cite{bessho2016topological},\cite{arvanitakis2021brane} These currents are called BFV(Batalin-Fradkin-Vilkovisky) current algebras because of their relationship with BFV theories \cite{batalin1977relativistic},\cite{batalin1983generalized}. There Courant algebroids(degree 2 dg symplectic manifolds) are generalized to degree $n$ dg symplectic manifolds. BFV current algebras and degree $n$ dg symplectic manifolds can be seen as a higher analog of the second line of ($\ref{2}$).

The aim of this paper is to give a higher analog of the third line and fourth line of ($\ref{2}$). In other words, we consider how to make higher Poisson vertex algebras, higher Courant-Dorfman algebras, higher Lie conformal algebras and higher weak Courant-Dorfman algebras which are generalizations of BFV current algebras and algebras of functions of degree $n$ dg symplectic manifolds. In particular, with higher Courant-Dorfman algebras and higher Poisson vertex algebras, we may be able to find and unify more general current algebras including the BFV current algebras, and use the techniques of Poisson vertex algebras in the higher setting.

In this paper, we give a higher analog of the relation between Poisson vertex algebras and Courant-Dorfman algebras. First, we define higher Courant-Dorfman algebras by taking algebraic structures of functions of degree $n$ dg symplectic manifolds. We give some examples, including ordinary Courant-Dorfman algebras and higher Dorfman bracket on $TM\oplus\wedge^{n-1}T^{*}M$. We also give an extended version of higher Courant-Dorfman algebras, whose definition is more natural when we consider the relation with higher PVAs.

Second, we check that non-degenerate higher Courant-Dorfman algebras have a similar property to the non-degenerate Courant-Dorfman algebras. In particular, we make a graded Poisson algebra of degree $-n$ from a non-degenerate higher Courant-Dorfman algebra. This graded Poisson algebra is a generalization of the graded Poisson algebra of degree $-2$ introduced in \cite{keller2015deformation},\cite{roytenberg2009courant}. For a non-degenerate Courant-Dorfman algebra from a finite-dimensional graded vector bundle, this graded Poisson algebra is isomorphic to the algebra of functions of dg symplectic manifolds.   

Third, we define a higher analog of Lie conformal algebras and Poisson vertex algebras, which are to higher Courant-Dorfman algebras what Poisson vertex algebras are to Courant-Dorfman algebras. We derive a weak notion of higher Courant-Dorfman algebras from higher Lie conformal algebras, and give the correspondence between higher Poisson vertex algebras and higher Courant-Dorfman algebras in Theorem 4.1. This correspondence is a higher generalization of the correspondence between Courant-Dorfman algebras and Poisson vertex algebras, and the main result of this paper.

\begin{theorem*}

There is a bijection between higher Poisson vertex algebras generated by elements of degree $0\leq i\leq n-1$ and extended higher Courant-Dorfman algebras.

\end{theorem*}

 Moreover, we check higher Lie conformal algebras and higher Poisson vertex algebras have LCA-like and PVA-like properties.  In particular, we show we can construct a graded Lie algebra out of the tensor product of a higher LCA and an arbitrary differential graded-commutative algebra (dgca for short) and a graded Poisson algebra out of the tensor product of a higher PVA and an arbitrary dgca. Taking a tensor product of the higher Courant-Dorfman algebra arising from a dg symplectic manifold of degree $n$ and de-Rham complex of a $n-1$ dimensional manifold, we see the associated Poisson algebras can be seen as an algebraic description of BFV current algebras. This is the higher generalization of Alekseev-Strobl Poisson vertex algebras.
 
 The higher generalization of ($\ref{2}$) are summarized as follows. 

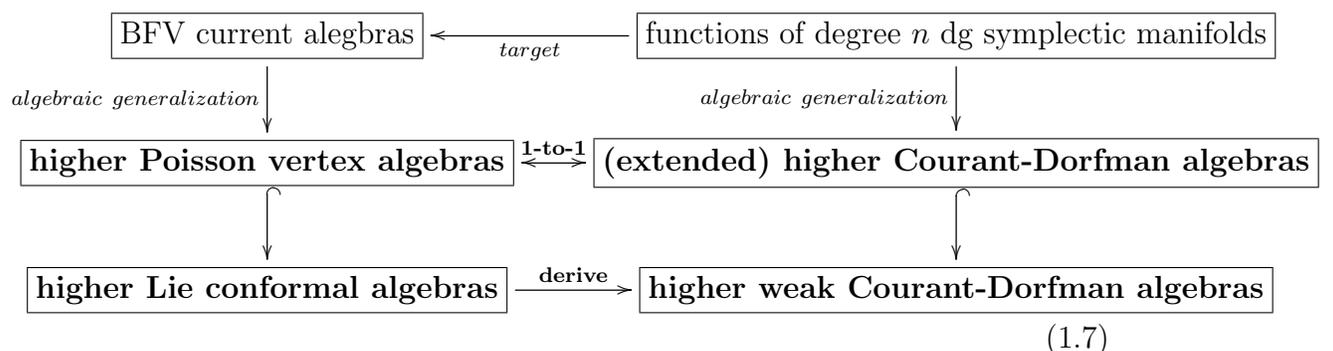
\begin{equation}
  \xymatrix{
    \fbox{BFV current alegbras} \ar@{^{(}->}[d] & \fbox{functions of degree $n$ dg symplectic manifolds}  \ar[l]^-{target} \ar@{^{(}->}[d] \\
    \fbox{\textbf{higher Poisson vertex algebras}} \ar@{<->}[r]^-{\textbf{1-to-1}}  \ar@{^{(}->}[d] & \fbox{\textbf{(extended) higher Courant-Dorfman algebras}} \ar@{^{(}->}[d]\\
    \fbox{\textbf{higher Lie conformal algebras}}  \ar[r]^-{\textbf{derive}}& \fbox{\textbf{higher weak Courant-Dorfman algebras}}
    }
\end{equation}

In the case of $n=2$, this coincides with ($\ref{2}$). The bold parts (second line and third line) are defined and studied in this paper.

\section{Definitions and examples of higher Courant-Dorfman algebras}\label{sec2}

In this section, we define higher Courant-Dorfman algebras of degree $n$ and give examples. The definition of these algebras of degree $2$ coincides with that of Courant-Dorfman algebras. 

Let $R=E^{0}$ be a commutative algebra over a ring $K\supset\mathbb{Q}$, and $E=\oplus_{1\leq i\leq n-1}E^{i}$ be a graded $R$-module, where $E^{i}$ has degree $i$. Define a pairing $\langle ,\rangle:E\otimes E\rightarrow R$ such that $\langle  a,b\rangle=0$ unless $|a|+|b|=n$. Consider the graded-commutative algebra freely generated by $E$ over $R$ and denote it by $\tilde{\mathcal{E}}=(\mathcal{E}^{k})_{k\in\mathbb{Z}}$. We restrict this graded-commutative algebra to the elements of degree $n-1\geq k\geq0$ and denote it by $\mathcal{E}=(\mathcal{E}^{k})_{n-1\geq k\geq0}$. The pairing $\langle ,\rangle$ can be extended to $\langle,\rangle:\mathcal{E}\otimes\mathcal{E}\rightarrow\mathcal{E}$ of degree $-n$ by the Leibniz rule
\begin{equation}
\langle  a,b\cdot c\rangle=\langle  a,b\rangle\cdot c+(-1)^{(|a|-n)|b|}b\cdot\langle  a,c\rangle,
\end{equation}
\begin{equation}
\langle  a\cdot b,c\rangle=a\cdot\langle  b,c\rangle+(-1)^{(|c|-n)|b|}\langle  a,c\rangle\cdot b.
\end{equation}

\begin{definition}

$\mathcal{E}=(\mathcal{E}^{k})_{n-1\geq k\geq0}$ is \textit{a higher Courant-Dorfman algebra} of degree $n$ if $\mathcal{E}$ has a differential $d:\mathcal{E}^{k}\rightarrow\mathcal{E}^{k+1}$ which satisfies $d^{2}=0$ and $d(a\cdot b)=(da)\cdot b+(-1)^{|a|}a\cdot (db)$ and a bracket $[,]:\mathcal{E}\otimes\mathcal{E}\rightarrow\mathcal{E}$ of degree $1-n$ which, together with the pairing $\langle ,\rangle:\mathcal{E}\otimes\mathcal{E}\rightarrow \mathcal{E}$ of degree $-n$, satisfy the conditions:

\begin{description}

\item[sesquilinearity]: 
\begin{equation}\label{ses}
\langle  da,b\rangle=-(-1)^{|a|-n}[a,b], [da,b]=0.
\end{equation}

\item[skew-symmetry]: 
\begin{equation}
\label{ss}
[a,b]+(-1)^{(|a|+1-n)(|b|+1-n)}[b,a]=(-1)^{|a|+1-n}d\langle  a,b\rangle,
\end{equation}
\begin{equation}\label{skew}
\langle  a,b\rangle=-(-1)^{(|a|-n)(|b|-n)}\langle  b,a\rangle.
\end{equation}

\item[Jacobi identity]: 
\begin{equation}
[a,[b,c]]=[[a,b],c]+(-1)^{(|a|+1-n)(|b|+1-n)}[b,[a,c]],
\end{equation}
\begin{equation}\label{cj}
[a,\langle  b,c\rangle]=\langle [a,b],c\rangle+(-1)^{(|a|+1-n)(|b|-n)}\langle  b,[a,c]\rangle,
\end{equation}
\begin{equation}
\langle  a,\langle  b,c\rangle\rangle=\langle \langle  a,b\rangle,c\rangle+(-1)^{(|a|-n)(|b|-n)}\langle  b,\langle  a,c\rangle\rangle.
\end{equation}

\item[Leibniz rule]:
\begin{equation}
[a,b\cdot c]=[a,b]\cdot c+(-1)^{(|a|+1-n)|b|}b\cdot[a,c].
\end{equation}

\end{description}

\end{definition}

Restricting the bracket to $\mathcal{E}^{n-1}\otimes\mathcal{E}^{n-1}\rightarrow\mathcal{E}^{n-1}$, it follows that $\mathcal{E}^{n-1}$ is a Leibniz algebra because of the Jacobi identity. However, it may fail to make $\mathcal{E}^{n-1}$ into a Lie algebra as skew-symmetry could fail due to the right-hand side of (\ref{ss}).

Next, we define the non-degeneracy condition, and fullness condition, like Courant-Dorfman algebras.   

\begin{definition}
The bilinear form $\langle ,\rangle$ gives rise to a map
\begin{equation}
(-)^{\flat}:E\rightarrow E^{\vee}=Hom_{R}(E,R)
\end{equation}
such that for all $1\leq i\leq n-1$ the restriction to $E^{i}$ satisfies
\begin{equation}
(-)^{\flat}|_{E^{i}}:E^{i}\rightarrow (E^{n-i})^{\vee}=Hom_{R}(E^{n-i},R),
\end{equation}
defined by
\begin{equation}
e^{\flat}(e')=\langle  e,e'\rangle.
\end{equation}
$\langle ,\rangle$ is non-degenerate if $(-)^{\flat}$ is an isomorphism, and a higher Courant-Dorfman algebra is non-degenerate if $\langle ,\rangle$ is non-degenerate. 
\end{definition}

When a higher Courant-Dorfman algebra is non-degenerate, the inverse map of $(-)^{\flat}$ is denoted by
\begin{equation}
(-)^{\sharp}:E^{\vee}\rightarrow E
\end{equation} 
and there is a graded-symmetric bilinear form
\begin{equation}
\{-,-\}:E^{\vee}\otimes_{R}E^{\vee}\rightarrow R
\end{equation}
defined by
\begin{equation}\label{nondeg}
\{\lambda,\mu\}=\langle \lambda^{\sharp},\mu^{\sharp}\rangle.
\end{equation}

\begin{definition}
$\langle ,\rangle$ is full if ,for every $1\leq i\leq n-1$, every $a\in R$ can be written as a finite sum $a=\sum_{j}\langle  x_{j},y_{j}\rangle$ with $x_{j}\in E^{i},y_{j}\in E^{n-i}$.

\end{definition}

Define the anchor map
\begin{equation}
\rho:E^{n-1}\rightarrow\mathfrak{X}=Der(R,R)
\end{equation}
by setting
\begin{equation}
\rho(e)\cdot f=\langle  e,df\rangle.
\end{equation}

We can define a Dirac submodule, like an ordinary Courant-Dorfman algebra. 

\begin{definition}

Suppose $\mathcal{E}$ is a higher Courant-Dorfman algebra. An $R$-submodule $\mathcal{D}\subset \mathcal{E}$ is said to be a Dirac submodule if $\mathcal{D}$ is isotropic with respect to $\langle ,\rangle$ and closed under $[-,-]$.
 
\end{definition}

We give some examples.

\begin{example}

Consider the case $n=2$. In this case, there is an $R$-module $E^{1}$, a pairing $\langle ,\rangle:E^{1}\otimes E^{1}\rightarrow R$, a derivation $d:R\rightarrow E^{1}$, and three brackets $[,]:R\otimes E^{1}\rightarrow R$,$[,]:E^{1}\otimes R\rightarrow R$,and $[,]:E^{1}\otimes E^{1}\rightarrow E^{1}$.

From the sesquilinearity, we get $[e,f]=\langle  e,df\rangle, [f,e]=-\langle  df,e\rangle$. For other operations, one can see that the above definition reduces to the definition of a Courant-Dorfman algebra. 

\end{example}

\begin{example}

Given a commutative algebra $R$, let $E^{n-1}=\mathfrak{X}=Der(R,R),E^{1}=\Omega^{1}(K\ddot{a}hler\ differential)$. In this case, $\mathcal{E}^{n-1}=\mathfrak{X}\oplus\Omega^{n-1}$. We define the pairing $\langle ,\rangle:E\otimes E\rightarrow R$ by $\langle v,\alpha\rangle=\alpha(v)$ for $v\in\mathfrak{X},\alpha\in\Omega^{1}$ and extend the pairing by the Leibniz rule. Explicitly the pairing  $\langle ,\rangle:\mathcal{E}\otimes\mathcal{E}\rightarrow \mathcal{E}$ of degree $-n$ becomes
\begin{equation}
\langle  v,\alpha\rangle=\iota_{v}\alpha, \langle v_{1},v_{2}\rangle=0, \langle \alpha_{1},\alpha_{2}\rangle=0
\end{equation}
where $v,v_{1},v_{2}\in\mathfrak{X}$, $\alpha,\alpha_{1},\alpha_{2}\in\Omega^{i}(1\leq i\leq n-1)$.

Then it becomes a higher Courant-Dorfman algebra with respect to
\begin{equation}
[v,\alpha]=L_{v}\alpha, [\alpha,v]=(-1)^{i+1-n}d(\iota_{v}\alpha)-L_{v}\alpha
\end{equation}
\begin{equation}
[v_{1},v_{2}]=[v_{1},v_{2}]_{\mathfrak{X}}+\iota_{v_{1}}\iota_{v_{2}}\omega\ (\omega\in\Omega^{n+1,cl}), [\alpha_{1},\alpha_{2}]=0
\end{equation}
where $v,v_{1},v_{2}\in\mathfrak{X}$, $\alpha,\alpha_{1},\alpha_{2}\in\Omega^{i}(1\leq i\leq n-1)$, and $d$ is the de-Rham differential on $\Omega^{i}$ and $[,]_{\mathfrak{X}}$ is the commutator of $Der(R,R)$. In the case of $R=C^{\infty}(M)$ for a smooth manifold $M$ and $\mathcal{E}^{n-1}=TM\oplus\wedge^{n-1}T^{*}M$, and the bracket $[,]$ is called a higher Dorfman bracket. In this case, the Dirac submodule is studied in \cite{bursztyn2019higher}.

\end{example}

\begin{example}

Let $(\mathcal{M},\omega,\Theta)$ be a degree $n$ dg symplectic manifold and $C=C^{n-1}(C^{\infty}(\mathcal{M}))=\{f\in C^{\infty}(\mathcal{M}):|f|\leq n-1\}$. This is a higher Courant-Dorfman algebra with 
\begin{equation}
[a,b]=\{\{a,\Theta\},b\},\ \langle  a,b\rangle=\{a,b\},\ da=\{\Theta,a\}.
\end{equation}
where $a,b\in C^{\infty}(\mathcal{M})$ and $\{-,-\}$ is the graded Poisson bracket induced from $\omega$.

In the previous example, the higher Courant-Dorfman algebra on $\mathcal{E}^{n-1}=TM\oplus\wedge^{n-1}T^{*}M$ coincides the algebra on $C=C^{n-1}(C^{\infty}(T^{*}[n]T[1]M))$.
\end{example}

\begin{example}

As a variant of Example 2.2, we can replace $\mathfrak{X}$ by a Lie-Rinehart algebra $(R,L)$ and let $E^{n-1}=L,E^{1}=\Omega^{1}$. In this case, $\mathcal{E}^{n-1}=L\oplus\Omega^{n-1}$. We define the pairing $\langle ,\rangle:E\otimes E\rightarrow R$ by $\langle v,\alpha\rangle=\alpha(\rho(v))$ for $v\in L,\alpha\in\Omega^{1}$ and extend the pairing by the Leibniz rule. Explicitly the pairing  $\langle ,\rangle:\mathcal{E}\otimes\mathcal{E}\rightarrow \mathcal{E}$ of degree $-n$ becomes
\begin{equation}
\langle  v,\alpha\rangle=\iota_{\rho(v)}\alpha, \langle v_{1},v_{2}\rangle=0, \langle \alpha_{1},\alpha_{2}\rangle=0
\end{equation}
where $v,v_{1},v_{2}\in\mathfrak{X}$, $\alpha,\alpha_{1},\alpha_{2}\in\Omega^{i}(1\leq i\leq n-1)$. Then it becomes a higher Courant-Dorfman algebra with respect to
\begin{equation}
[v,\alpha]=L_{\rho(v)}\alpha, [\alpha,v]=(-1)^{i+1-n}d(\iota_{\rho(v)}\alpha)-L_{\rho(v)}\alpha, 
\end{equation}
\begin{equation}
[v_{1},v_{2}]=[v_{1},v_{2}]_{L}+\iota_{(\rho(v_{1}))}\iota_{(\rho(v_{2}))}\omega\ (\omega\in\Omega^{n+1,cl})
\end{equation}
for $v,v_{1},v_{2}\in L$ and $\alpha\in\Omega^{i}(1\leq i\leq n)$,where $d$ is the de-Rham differential on $\Omega^{i}$ and $[,]_{L}$ is the Lie bracket of the Lie-Rinehart algebra $(R,L)$.
\end{example}

In order to focus on the relation with higher Poisson vertex algebras, we should define extended higher Courant-Dorfman algebras, relaxing the condition on $\langle ,\rangle$.

\begin{definition}
Let $R=E^{0}$ be a commutative algebra, and $E=E^{i}(1\leq i\leq n-1)$ be a graded $R$-module. Consider the graded-commutative algebra freely generated by $E$ and denote it by $\tilde{\mathcal{E}}=(\mathcal{E}^{k})_{k\in\mathbb{Z}}$. We restrict this graded-commutative algebra to the elements of degree $n-1\geq k\geq0$ and denote it by $\mathcal{E}=(\mathcal{E}^{k})_{n-1\geq k\geq0}$.

$\mathcal{E}=(\mathcal{E}^{k})_{n-1\geq k\geq0}$ is \textit{an extended higher Courant-Dorfman algebra} of degree $n$ if $\mathcal{E}$ has a differential $d:\mathcal{E}^{k}\rightarrow\mathcal{E}^{k+1}$ which  satisfies $d^{2}=0$ and $d(a\cdot b)=(da)\cdot b+(-1)^{|a|}a\cdot (db)$, a pairing $\langle ,\rangle :\mathcal{E}\otimes\mathcal{E}\rightarrow\mathcal{E}$ of degree $-n$ and a bracket $[,]:\mathcal{E}\otimes\mathcal{E}\rightarrow\mathcal{E}$ of degree $1-n$ which satisfies the sesquilinearity, skew-symmetry, Jacobi identity, and Leibniz rule.

\end{definition}

The difference with a higher Courant-Dorfman algebra is that an extended Courant-Dorfman algebra allows the pairing $\langle,\rangle:E^{i}\otimes E^{j}\rightarrow E^{i+j-n}$ with $i+j\geq n+1$. From the viewpoint of graded geometry, these algebras include the case that the base manifold is a graded manifold.

\section{Non-degenerate higher Courant-Dorfman algebras and degree $n$ dg symplectic manifolds}\label{sec3}

In this section, we consider the case that  $\langle ,\rangle$ is non-degenerate, and study the relationship between the algebras and functions of degree $n$ dg symplectic manifolds. We construct two equivalent graded Poisson algebra of degree $-n$, generalizing the Keller-Waldmann Poisson algebras\cite{keller2015deformation},which can be seen as an algebralization of functions of degree $n$ dg symplectic manifolds. We assume that each $E^{i}$ is a projective, finitely generated module over $R$, and that $\langle ,\rangle $ is non-degenerate and full.

\begin{definition}
We assume $r\geq n$. 
\begin{equation}
\mathcal{C}^{r}(\mathcal{E})\subset\oplus_{1\leq j\leq n-1}\oplus_{1\leq k\leq r-j}\oplus_{\sum^{k}_{t=1}i_{t}=r-j}\mathrm{Hom}_{K}(E^{n-i_{1}}\otimes\cdots\otimes E^{n-i_{k}},E^{j})
\end{equation}
consists of elements $C$ for which there exists a $K$-multilinear map
\begin{equation}
\sigma_{C}\in \oplus_{0\leq l\leq r-n}\oplus_{\sum^{l}_{t'=1} i_{t'}=r-n}\mathrm{Hom}_{K}(E^{n-i_{1}}\otimes\cdots\otimes E^{n-i_{l}},\mathfrak{X}),
\end{equation}
satisfying the following conditions:

(1)For all $u,w,x_{1},...,x_{l}\in E$, we have
\begin{equation}
\sigma_{C}\langle  u,w\rangle=\langle  C(u),w\rangle+\langle  u,C(w)\rangle,
\end{equation}
\begin{equation}\label{as}
\sigma_{C}(x_{1},...,x_{l})\langle  u,w\rangle=\langle  C(x_{1},...,x_{l},u),w\rangle+\langle  u,C(x_{1},...,x_{l},w)\rangle.
\end{equation}

(2)For all $x_{1},...,x_{k},u\in E$, we have, for all $1\leq i\leq k-1$,
\begin{align}\label{symbol2}
&\langle  C(x_{1},...x_{i},x_{i+1},...,x_{k})-(-1)^{(|x_{i}|-n)(|x_{i+1}|-n)}C(x_{1},....,x_{i+1},x_{i},...,x_{k}),u\rangle \notag \\
&=\sigma_{C}(x_{1},...,x_{i-1},x_{i+2},....,x_{k},u)\langle  x_{i},x_{i+1}\rangle.
\end{align}

Furthermore, $\mathcal{C}^{0}(\mathcal{E})=R, \mathcal{C}^{i}(\mathcal{E})=\mathcal{E}^{i}$ for $1\leq i\leq n-1$ and define
\begin{equation}
\mathcal{C}^{\bullet}(\mathcal{E})=\oplus_{r\geq0}\mathcal{C}^{r}(\mathcal{E}).
\end{equation}

When $C\in\mathcal{C}^{r}(\mathcal{E})$, we denote $|C|:=r$. Note that $\sigma_{C}$ is defined only for $r\geq n$. We call $\sigma_{C}$ the symbol of $C$. 

\end{definition}

Given $C\in\mathcal{C}^{r}(E)$, we can get the map $C(x_{1},...,x_{l},-):E\rightarrow E$ of degree $c:=r-n(l+1)+|x_{1}|+|x_{2}|+\cdots+|x_{l}|$. For the right side of (\ref{as}), $c=0$ due to the condtions of $\sigma_{C}$.

In particular, if $D\in\mathcal{C}^{r}(E)$ for $r<2n$, we can get the map $D(-):E\rightarrow E$of degree $d:=r-n$. For $x\in\ E^{i}$ and $C\in\mathcal{C}^{r}(\mathcal{E})$, we define $\iota_{x}C\in\mathcal{C}^{r+i-n}(\mathcal{E})$ by
\begin{equation}
\iota_{x}C(x_{1},...,x_{t})=C(x,x_{1},...,x_{t}).
\end{equation}

Define 
\begin{equation}
d_{C}\in\oplus_{1\leq l\leq r-n-k}\oplus_{\sum^{l}_{t'=1} i_{t'}=r-n-k}\mathrm{Hom}_{K}(E^{n-i_{1}}\otimes\cdots\otimes E^{n-i_{l}},\mathfrak{X}\otimes E^{k})
\end{equation}
by
\begin{equation}
\langle d_{C}(x_{1},...,x_{l})a,y\rangle:=\sigma_{C}(x_{1},...,x_{l},y)a.
\end{equation}

Comparing to (\ref{symbol2}), we get
\begin{align}
& C(x_{1},...x_{i},x_{i+1},...,x_{k})-(-1)^{(|x_{i}|-n)(|x_{i+1}|-n)}C(x_{1},....,x_{i+1},x_{i},...,x_{k}) \notag \\
&=d_{C}(x_{1},...,x_{i-1},x_{i+2},....,x_{k})\langle  x_{i},x_{i+1}\rangle.
\end{align}
for all  $x_{1},...,x_{k},u\in E$, $1\leq i\leq k-1$.
For $a\in R$ and $C\in\mathcal{C}^{r}(\mathcal{E})$, we define the map $d_{C}a\in \mathcal{C}^{r-n}(\mathcal{E})$ as $d_{C}a(x_{1},...,x_{l})=d_{C}(x_{1},...,x_{l})a$. Then we have
\begin{equation}
\iota_{x_{i}}\iota_{x_{i+1}}C-(-1)^{(|x_{i}|-n)(|x_{i+1}|-n)}\iota_{x_{i+1}}\iota_{x_{i}}C=d_{C}\langle x_{i},x_{i+1}\rangle.
\end{equation}

We can use instead of elements $C\in\mathcal{C}^{r}(\mathcal{E})$ $K$-multilinear forms 
\begin{equation}
\omega\in \oplus_{1\leq k\leq r}\oplus_{\sum^{k}_{t=1}i_{t}=r}\mathrm{Hom}_{K}(E^{n-i_{1}}\otimes\cdots\otimes E^{n-i_{k}},R)
\end{equation}
defined by $\omega(x_{1},...,x_{t})=\langle C(x_{1},...,x_{t-1}),x_{t}\rangle $.

\begin{definition}
For $r\geq1$ the subspace 
\begin{equation}
\Omega^{r}_{\mathcal{C}}(\mathcal{E})\subset\oplus_{1\leq k\leq r}\oplus_{\sum^{k}_{t=1}i_{t}=r}\mathrm{Hom}_{K}(E^{n-i_{1}}\otimes\cdots\otimes E^{n-i_{k}},R)
\end{equation}
consists of elements $\omega$ satisfying the following conditions;

(1)
\begin{equation}
\omega(x_{1},...,ax_{k})=a\omega(x_{1},...,x_{k}),
\end{equation}
for all $a\in R$.

(2)For $r\geq2$, there exists a multilinear map,

\begin{equation}
\sigma_{\omega}\in \oplus_{1\leq l\leq r-n}\oplus_{\sum^{l}_{t'=1} i_{t'}=r-n}\mathrm{Hom}_{K}(E^{n-i_{1}}\otimes\cdots\otimes E^{n-i_{l}},\mathfrak{X}),
\end{equation}
such that, for all $1\leq i\leq k-1$,

\begin{align}
&\omega(x_{1},...x_{i},x_{i+1},...,x_{k})-(-1)^{(|x_{i}|-n)(|x_{i+1}|-n)}\omega(x_{1},....,x_{i+1},x_{i},...,x_{k}) \notag \\
&=\sigma_{\omega}(x_{1},...^{\wedge^{i}}...^{\wedge^{i+1}},x_{k})\langle  x_{i},x_{i+1}\rangle.
\end{align}

\end{definition}

By the non-degeneracy of $\langle,\rangle$, we get the following Lemma.

\begin{lemma}
There is an isomorphism of graded $R$-modules
\begin{equation}
\mathcal{C}^{\bullet}(\mathcal{E})\rightarrow \Omega^{\bullet}_{\mathcal{C}}(\mathcal{E}),
\end{equation}
given by
\begin{equation}
\omega(x_{1},...,x_{t})=\langle  C(x_{1},...,x_{t-1}),x_{t}\rangle.
\end{equation}

\end{lemma}

Next, we define the graded Lie bracket on $\mathcal{C}^{\bullet}(\mathcal{E})$.

\begin{proposition}
The map
\begin{equation}
[,]:\mathcal{C}^{r}(\mathcal{E})\otimes\mathcal{C}^{s}(\mathcal{E})\rightarrow\mathcal{C}^{r+s-n}(\mathcal{E}),
\end{equation}
defined by
\begin{equation}
\label{l1}
[a,b]=0, [a,x]=0=[x,a], [x,y]=\langle  x,y\rangle, [D,a]=\sigma_{D}a=-[a,D],
\end{equation}
\begin{equation}
\label{l2}
[C,x]=\iota_{x}C=-(-1)^{(r+n)(|x|+n)}[x,C],\ [C,a]=d_{C}a=-(-1)^{(r+n)n}[a,C]
\end{equation}
for elements $a,b\in R, x,y\in\mathcal{C}^{s}(\mathcal{E})$ for $s<n$, $D\in\mathcal{C}^{n}(\mathcal{E}),C\in\mathcal{C}^{r}(\mathcal{E})$ for $r\geq n$, and by the recursion,
\begin{equation}
\label{r}
\iota_{x}[C_{1},C_{2}]=[[C_{1},C_{2}],x]=[C_{1},[C_{2},x]]-(-1)^{(|C_{1}|+n)(|C_{2}|+n)}[C_{2},[C_{1},x]],
\end{equation}
is well-defined and makes $\mathcal{C}^{\bullet}(\mathcal{E})$ a graded Lie algebra of degree $-n$. 
\end{proposition}

\begin{proof}

It suffices to show that the recursion ($\ref{r}$) is consistent with ($\ref{l1}$) and ($\ref{l2}$), that $[C_{1},C_{2}]\in\mathcal{C}^{r+s-n}(\mathcal{E})$, and that the bracket satisfies the conditions for a graded Lie algebra. We use the formula $[D,x]=D(x)$ which can be obtained as a consequence of  ($\ref{l2}$). 

The consistency can be checked as follows:

\begin{align}\label{start}
[[D,x],y]&=\langle  D(x),y\rangle=(-1)^{(|x|-n)(|y|-n)}\langle  D(y),x\rangle+\sigma_{D}\langle  x,y\rangle \notag \\
&=(-1)^{(|x|-n)(|y|-n)}[[D,y],x]+[D,[x,y]].
\end{align}
\begin{align}
[[C,x],y]&=\iota_{y}\iota_{x}C=(-1)^{(|x|-n)(|y|-n)}\iota_{x}\iota_{y}C+d_{C}\langle  x,y\rangle \notag \\
&=(-1)^{(|x|-n)(|y|-n)}[[C,y],x]+[C,[x,y]].
\end{align}

Next, we check that $[C_{1},C_{2}]$ is an element in $\mathcal{C}^{r+s-n}(\mathcal{E})$. Let $N:=|C_{1}|+|C_{2}|$. For $N\leq 2n-1$, the claim is clear. For $N=2n$, we consider three cases. If $a\in R$ and $C\in\mathcal{C}^{2n}(\mathcal{E})$, then $[C,a]=d_{C}a\in\mathcal{C}^{n}(\mathcal{E})$ and
\begin{equation}
[[C,a],b]=[C,[a,b]]-(-1)^{n}[a,[C,b]].
\end{equation}  
If $x\in E^{i}$ and $C\in\mathcal{C}^{2n-i}(\mathcal{E})$, then $[C,x]=\iota_{C}x\in\mathcal{C}^{n}(\mathcal{E})$ and
\begin{equation}
[[C,x],a]=[C,[x,a]]-(-1)^{(n-i)(i+n)}[x,[C,a]].
\end{equation}
If $D_{1},D_{2}\in\mathcal{C}^{n}(\mathcal{E})$, then $[D_{1},D_{2}]\in\mathcal{C}^{n}(\mathcal{E})$ with
\begin{equation}
\sigma_{[D_{1},D_{2}]}a=\sigma_{D_{1}}\sigma_{D_{2}}a-\sigma_{D_{2}}\sigma_{D_{1}}a,
\end{equation}
\begin{equation}\label{end}
[[D_{1},D_{2}],a]=[D_{1},[D_{2},a]]-[D_{2},[D_{1},a]].
\end{equation}

Let $C_{1}\in\mathcal{C}^{r}(\mathcal{E}),C_{2}\in\mathcal{C}^{s}(\mathcal{E})$ with $r+s\geq2n+1$. Consider a map $h:R\rightarrow\mathcal{C}^{r+s-2n}(\mathcal{E})$ defined by
\begin{equation}
h(a)=[C_{1},[C_{2},a]]-(-1)^{(r+n)(s+n)}[C_{2},[C_{1},a]].
\end{equation}
Then $[C_{1},C_{2}]\in\mathcal{C}^{r+s-n}(\mathcal{E})$ and the symbol is
\begin{equation}
\sigma_{[C_{1},C_{2}]}(x_{1},...,x_{t})a=\langle  h(a)(x_{1},...,x_{t-1}),x_{t}\rangle.
\end{equation}

The skew-symmetry is clear by the construction. We check the Jacobi identity. It suffices to show
\begin{equation}
J(C_{1},C_{2},C_{3}):=[[C_{1},C_{2}],C_{3}]-[C_{1},[C_{2},C_{3}]]+(-1)^{(|C_{1}|+n)(|C_{2}|+n)}[C_{2},[C_{1},C_{3}]]=0.
\end{equation}

We prove the claim by induction for $N:=|C_{1}|+|C_{2}|+|C_{3}|$. For $1\leq N< 2n$, it is clear because the degree of $[,]$ is $-n$. For $N=2n$, it follows from (\ref{start})-(\ref{end}).  By the recursion rule,
\begin{align}
&[J(C_{1},C_{2},C_{3}),x] \notag \\
=&(-1)^{(|C_{2}|-n)(|x|-n)+(|C_{3}|-n)(|x|-n)}J([C_{1},x],C_{2},C_{3}) \notag \\
+&(-1)^{(|C_{3}|-n)(|x|-n)}J(C_{1},[C_{2},x],C_{3})+J(C_{1},C_{2},[C_{3},x]).
\end{align}
Each $C_{i}$ satisfies $|[C_{i},x]|<|C_{i}|$, and by induction we can obtain $[J(C_{1},C_{2},C_{3}),x]=0$ for all $x\in E$. 

For $N\geq2n+1$, $|J(C_{1},C_{2},C_{3})|\geq1$, so there is $x\in E$ such that $[J(C_{1},C_{2},C_{3}),x]\neq 0$ unless $J(C_{1},C_{2},C_{3})=0$. Therefore we conclude $J(C_{1},C_{2},C_{3})=0$ for all $C_{1},C_{2},C_{3}\in\mathcal{C}^{\bullet}(E)$.
\end{proof}

\begin{proposition}
There exists an associative, graded-commutative $K$-bilinear product $\wedge$ of degree 0 on $\mathcal{C}^{\bullet}(\mathcal{E})$ uniquely defined by
\begin{equation}
a\wedge b=ab=b\wedge a, a\wedge x=ax=x\wedge a,
\end{equation}
for $a,b\in R$ and $x\in E$ and by the recursion rule
\begin{equation}
[C_{1}\wedge C_{2},x]=(-1)^{rs}C_{2}\wedge[C_{1},x]+C_{1}\wedge[C_{2},x]
\end{equation}
for $C_{1}\in\mathcal{C}^{r}(E), C_{2}\in\mathcal{C}^{s}(E)$.

\end{proposition}

\begin{proof}
It suffices to show that
\begin{equation}
C_{1}\wedge C_{2}:(x_{1},...,x_{t})\mapsto[C_{1}\wedge C_{2},x_{1}](x_{2},...,x_{t})
\end{equation}
is an element in $\mathcal{C}^{r+s}(\mathcal{E})$ for $N=r+s\geq n$. Consider the map
\begin{equation}
h(a)=(-1)^{rs}C_{2}\wedge[C_{1},a]+C_{1}\wedge[C_{2},a].
\end{equation}
Then $C_{1}\wedge C_{2}\in\mathcal{C}^{r+s}(\mathcal{E})$ and the symbol is
\begin{equation}
\sigma_{[C_{1},C_{2}]}(x_{1},...,x_{t})a=\langle  h(a)(x_{1},...,x_{t-1}),x_{t}\rangle.
\end{equation}

\end{proof}

\begin{theorem}

 $(\mathcal{C}^{\bullet}(\mathcal{E}),[,],\wedge)$ is a graded Poisson algebra of degree $-n$.

\end{theorem}

\begin{proof}

It suffices to show the Leibniz rule
\begin{equation}
[C_{1}\wedge C_{2},C_{3}]=(-1)^{rs}C_{2}\wedge[C_{1},C_{3}]+C_{1}\wedge[C_{2},C_{3}].
\end{equation}

We can check by direct calculations and induction over $N:=|C_{1}|+|C_{2}|+|C_{3}|$.

\end{proof}

Since $\mathcal{C}^{\bullet}(\mathcal{E})\simeq \Omega^{\bullet}_{\mathcal{C}}(\mathcal{E})$, we can define a graded Poisson algebraic structure on $\Omega^{\bullet}_{\mathcal{C}}(\mathcal{E})$. This bracket is an extension of the bracket $\{-,-\}:E^{\vee}\otimes_{R}E^{\vee}\rightarrow R$ defined in (\ref{nondeg}).

We can construct $\omega_{\phi}\in\Omega^{r}_{C}(\mathcal{E})\simeq\mathcal{C}^{r}(\mathcal{E})$ from the map $\phi:\mathcal{E}^{i_{1}}\otimes\mathcal{E}^{i_{2}}\otimes\cdots\otimes\mathcal{E}^{i_{m}}\rightarrow\mathcal{E}^{i_{1}+\cdots+i_{m}-mn+r}$ by
\begin{equation}\label{diff}
\omega_{\phi}(e_{1},e_{2},...,e_{k})=\langle \cdots\langle \phi(e_{1},...,e_{m}),e_{m+1}\rangle\cdots\rangle,e_{k}\rangle
\end{equation}
where $e_{t}\in E^{i_{t}}$ and $\sum^{k}_{t=1} i_{t}=kn-r$. When $k=m$, we define $\omega_{\phi}(e_{1},e_{2},...,e_{m})=\phi(e_{1},...,e_{m})$.

When $\phi:\mathcal{E}^{i_{1}}\otimes\mathcal{E}^{i_{2}}\rightarrow\mathcal{E}^{i_{1}+i_{2}-2n+r}$ and $\phi':\mathcal{E}^{j_{1}}\otimes\mathcal{E}^{j_{2}}\rightarrow\mathcal{E}^{j_{1}+j_{2}-2n+r'}$ are binary brackets, the Lie bracket is
\begin{align}
& [\omega_{\phi},\omega_{\phi'}](e_{1},e_{2},...,e_{k}) \notag \\
&=\langle \cdots\langle \phi(e_{1},\phi'(e_{2},e_{3}))-\phi(\phi'(e_{1},e_{2}),e_{3})-(-1)^{(|e_{1}|+r)(|e_{2}|+r')}\phi(e_{2},\phi'(e_{1},e_{3})),e_{4}\rangle\cdots\rangle,e_{k}\rangle \notag \\
&+(-1)^{(r-n)(r-n')}\langle \cdots\langle \phi'(e_{1},\phi(e_{2},e_{3}))-\phi'(\phi(e_{1},e_{2}),e_{3})-(-1)^{(|e_{1}|+r)(|e_{2}|+r')}\phi'(e_{2},\phi(e_{1},e_{3})),e_{4}\rangle\cdots\rangle,e_{k}\rangle.
\end{align} 

Let $\phi$ be the bracket of the higher Courant-Dorfman algebra $[,]:\mathcal{E}\otimes\mathcal{E}\rightarrow\mathcal{E}$ of degree $1-n$. Then, $\omega_{\phi}$ satisfies $|\omega_{\phi}|=n+1$ and $[\omega_{\phi},\omega_{\phi}]=0$ as
\begin{align*}
[\omega_{\phi},\omega_{\phi}](e_{1},e_{2},...,e_{k})&=2\langle \cdots\langle [e_{1},[e_{2},e_{3}]]-[[e_{1},e_{2}],e_{3}]-(-1)^{(|e_{1}|+1-n)(|e_{2}|+1-n)}[e_{2},[e_{1},e_{3}]],e_{4}\rangle\cdots\rangle,e_{k}\rangle\\
&=0
\end{align*}

Therefore, the map $[\omega_{\phi},-]$ is degree 1 and squares to 0, so it defines a differential on $\mathcal{C}^{\bullet}(\mathcal{E})$. This can be seen as a higher derived bracket \cite{voronov2005higher} of this algebra.

Next, we define another Poisson algebra $\mathcal{R}^{\bullet}(\mathcal{E})$ generalizing the Rothstein algebra. This algebra is more useful for seeing the relation to the algebras of functions of graded symplectic manifolds.

\begin{definition}

Let $\mathfrak{X}:=Der(R)$. A connection $\nabla$ for the graded module $E=(E^{i})$ is a map $\nabla:\mathfrak{X}\times E\rightarrow E$ of degree 0 such that
\begin{equation}
\nabla_{aD}x=a\nabla_{D}x,
\end{equation}
\begin{equation}
\nabla_{D}(ax)=a\nabla_{D}x+D(a)x,
\end{equation}
for all $a\in R$, $x\in E$ and $D\in\mathfrak{X}$. If $\langle ,\rangle :E\otimes E\rightarrow R$ is an $R$-bilinear form, then $\nabla$ is called metric if in addition
\begin{equation}
D\langle x,y\rangle =\langle \nabla_{D}x,y\rangle+\langle  x,\nabla_{D}y\rangle,
\end{equation}
for all $x,y\in E$ and $D\in\mathfrak{X}$.

\end{definition}

If each $E^{i}$ is finitely generated and projective then it allows for a connection $\nabla$. If the $R$-bilinear form $\langle ,\rangle :E\otimes E \rightarrow R$ is non-degenerate and satisfies the skew-symmetry (\ref{skew}), then $\nabla$ can be chosen to be a metric connection. Indeed, if $\tilde{\nabla}$ is any connection and $\langle ,\rangle$ is non-degenerate then $\nabla$ defined by
\begin{equation}
\langle \nabla_{D}x,y\rangle=\frac{1}{2}(\langle \tilde{\nabla}_{D}x,y\rangle-\langle x,\tilde{\nabla}_{D}y\rangle+D\langle  x,y\rangle)
\end{equation}
is a metric connection.

\begin{definition}
The higher Rothstein algebra is defined as a symmetric algebra in the graded-commutative sense by
\begin{equation}
\mathcal{R}^{\bullet}(\mathcal{E})=\mathrm{Sym}(\oplus_{1\leq i\leq n-1}E^{i}[-i]\oplus\mathfrak{X}[-n]).
\end{equation}
\end{definition}

Next we introduce the curvature of $\nabla$. A given connection for $E$ extends to $\mathrm{Sym}(E)$ by imposing the Leibniz rule. Thus we can consider
\begin{equation}
R(D_{1},D_{2})\xi:=\nabla_{D_{1}}\nabla_{D_{2}}\xi-\nabla_{D_{2}}\nabla_{D_{1}}\xi-\nabla_{[D_{1},D_{2}]}\xi,
\end{equation} 
for $D_{i}\in\mathfrak{X}$ and $\xi\in \mathrm{Sym}(E)$. It defines an element
\begin{equation}
R(D_{1},D_{2})\in \mathrm{End}(\mathrm{Sym}(E)).
\end{equation}

When $\langle-,-\rangle$ is non-degenerate $R$-bilinear form and satisfies the skew-symmetry (\ref{skew}) and $\nabla$ is a metric connection, restricting $R(D_{1},D_{2})$ to $E$ gives a map $R(D_{1},D_{2}):E\rightarrow E$. For $x\in E^{i}$ and $y\in E^{n-i}$,
\begin{equation}
\langle R(D_{1},D_{2})x,y\rangle=(-1)^{i(n-i)}\langle R(D_{1},D_{2})y,x\rangle.
\end{equation}

$E^{i}$ is projective and finitely generated, so using the non-degenerate inner product $\langle,\rangle$ on $E$ we can define $r(D_{1},D_{2})\in Sym^{2}E$ of degree $n$ by
\begin{equation}
R(D_{1},D_{2})x=\langle  r(D_{1},D_{2}),x\rangle.
\end{equation}

With this preparation a Poisson structure can now be defined.

\begin{theorem}
Let $E$ be a graded module over $R$ endowed with a non-degenerate $R$-bilinear form $\langle-,-\rangle$ which satisfies the skew-symmetry (\ref{skew}) and $\nabla$ be a metric connection on $E$. Then there exists a unique graded Poisson structure $\{-,-\}_{R}$ on $\mathcal{R}^{\bullet}(\mathcal{E})$ of degree $-n$ such that
\begin{align}
\{a,b\}_{R}&=0=\{a,x\}_{R},\\
\{x,y\}_{R}&=\langle x,y\rangle=-(-1)^{(|x|-n)(|y|-n)}\{y,x\}_{R}, \\
\{D,a\}_{R}&=-D(a)=-\{a,D\}_{R},\\
\{D,x\}_{R}&=-\nabla_{D}x=-\{x,D\}_{R},\\
\{D_{1},D_{2}\}_{R}&=-[D_{1},D_{2}]+r(D_{1},D_{2})=-\{D_{2},D_{1}\}_{R},
\end{align}
for $a,b\in R$, $x,y\in E$ and $D_{1},D_{2}\in\mathfrak{X}$.

\end{theorem}

\begin{proof}

We can extend the bracket $\{\}_{R}$ to $\mathcal{R}(\mathcal{E})$ by the Leibniz rule from the above definition. The skew-symmetry is clear by construction.

Jacobi identity follows from the following identity
\begin{align}
\ &\nabla_{D_{1}}r(D_{2},D_{3})+\nabla_{D_{2}}r(D_{3},D_{1})+\nabla_{D_{3}}r(D_{1},D_{2}) \notag \\
+&r(D_{1},[D_{2},D_{3}])+r(D_{2},[D_{3},D_{1}])+r(D_{3},[D_{1},D_{2}])=0,
\end{align}
obtained from the Bianchi identity of the curvature $R(D_{1},D_{2})$.

\end{proof}

Next, we find the relation between $\mathcal{R}^{\bullet}(\mathcal{E})$ and  $\mathcal{C}^{\bullet}(\mathcal{E})$. Note that $\nabla_{D}\in\mathcal{C}^{n}(\mathcal{E})$ with $\sigma_{\nabla_{D}}=D$, which motivates the following definition.

\begin{definition}
Let the $R$-linear map $\mathcal{J}:\mathcal{R}^{\bullet}(\mathcal{E})\rightarrow\mathcal{C}^{\bullet}(\mathcal{E})$ be defined by
\begin{equation}
\mathcal{J}(a)=a,\mathcal{J}(x)=x,\mathcal{J}(D)=-\nabla_{D}
\end{equation}
for $a\in R,x\in E$ and $D\in\mathfrak{X}$ and extend by the Leibniz rule.

\end{definition}

\begin{proposition}
(1) The map $\mathcal{J}$ is a homomorphism of Poisson algebras.

(2) Let  $\phi\in\mathcal{R}^{\bullet}(\mathcal{E})$ with $r\geq n$, then
\begin{equation}
\mathcal{J}(\phi)(x_{1},...,x_{k})=\{\{...\{\phi,x_{1}\}_{R},...\}_{R},x_{k}\}_{R},
\end{equation}
and
\begin{equation}
\sigma_{\mathcal{J}(\phi)}(x_{1},...,x_{k-1})a=\{\{...\{\phi,x_{1}\}_{R},...\}_{R},x_{k-1}\}_{R},a\}_{R},
\end{equation}
for all $x_{i}\in E$ and $a\in R$.

\end{proposition}

\begin{proof}
(1) From the definition this is obvious for generators and it is true for all $\mathcal{R}^{\bullet}(\mathcal{E})$ by the Leibniz rule. 

(2) We can check $\mathcal{J}(\phi)(x)=[\mathcal{J}(\phi),x]=[\mathcal{J}(\phi),\mathcal{J}(x)]=\mathcal{J}(\{\phi,x\}_{R})$ with $\sigma_{\mathcal{J}(\phi)}a=\{\phi,a\}_{R}$, and show the claim by induction for $k$. Indeed, if $[\cdots[\mathcal{J}(\phi),x_{1}],\cdots],x_{k-1}]=\mathcal{J}\{\{...\{\phi,x_{1}\}_{R},...\}_{R},x_{k-1}\}_{R}$, then we compute 
\begin{align*}
\mathcal{J}(\phi)(x_{1},...,x_{k})&=[[\cdots[\mathcal{J}(\phi),x_{1}],\cdots],x_{k-1}],x_{k}]\\
&=[\mathcal{J}(\{\{...\{\phi,x_{1}\}_{R},...\}_{R},x_{k-1}\}_{R}),\mathcal{J}(x_{k})]\\
&=\mathcal{J}(\{\{...\{\phi,x_{1}\}_{R},...\}_{R},x_{k}\}_{R}).
\end{align*}
When $|\mathcal{J}(\phi)(x_{1},...,x_{k})|<n$, this is equal to $\{\{...\{\phi,x_{1}\}_{R},...\}_{R},x_{k}\}_{R}$.
\end{proof}

\begin{lemma}
Let  $\phi\in\mathcal{R}^{r}(\mathcal{E})$ with $r\geq 1$. Then
\begin{equation}
\mathcal{J}(\phi)(x_{1},...,x_{k})=\{\{...\{\phi,x_{1}\}_{R},...\}_{R},x_{k}\}_{R}=0,
\end{equation}
if and only if $\phi=0$.
\end{lemma}

\begin{proof}
It is true for $r=1,...,n-1$ due to the non-degeneracy of $\langle ,\rangle$. Suppose it is true for $1,...,r-1$ for some $r\geq n$. For $\phi\in\mathcal{R}^{r}(\mathcal{E})$, we have $|\{\phi,x\}_{R}|<r$. Therefore $\{\phi,x\}_{R}$ satisfies the condition if and only if $\{\phi,x\}_{R}=0$. Such $\phi$ satisfies
\begin{equation}
\{\phi,\langle  x,y\rangle\}=\{\phi,\{x,y\}\}=\{\{\phi,x\},y\}+(-1)^{(\phi+n)(|x|+n)}\{x,\{\phi,y\}\}=0,
\end{equation}
and $\{\phi,a\}_{R}=0$ due to the fullness of $\{,\}$. These equations imply that $\phi=0$.
\end{proof}

\begin{corollary}
Let $\hat{\mathcal{C}}^{\bullet}(\mathcal{E})$ be the subalgebra of $\mathcal{C}^{\bullet}(\mathcal{E})$ generated by $R,E$ and $\mathcal{C}^{n}(\mathcal{E})$. Then $\hat{\mathcal{C}}^{\bullet}(\mathcal{E})$ is closed under the bracket $[,]$ and $\mathcal{J}$ is an isomorphism of Poisson algebras
\begin{equation}
\mathcal{J}:\mathcal{R}^{\bullet}(\mathcal{E})\rightarrow\hat{\mathcal{C}}^{\bullet}(\mathcal{E}).
\end{equation}
\end{corollary}

\begin{proof}
$\mathcal{J}$ is injective due to the Lemma 3.2. From the Leibniz identity of $C\in\hat{\mathcal{C}}^{\bullet}(\mathcal{E})$ it follows that $\hat{\mathcal{C}}^{\bullet}(\mathcal{E})$ is the Poisson subalgebra of $\mathcal{C}^{\bullet}(\mathcal{E})$. If $D\in\mathcal{C}^{n}(\mathcal{E})$ we can define an element $\xi\in Sym(E)$ of degree $n$ by $\langle \xi,x\rangle=D(x)-\nabla_{\sigma_{D}}x$. It follows that $\{\xi-\sigma_{D},x\}_{R}=D(x)$, so $D\in\mathcal{J}(\mathcal{R}^{n}(\mathcal{E}))$, therefore $\mathcal{C}^{n}(\mathcal{E})\simeq\mathcal{R}^{n}(\mathcal{E})$.
\end{proof}

\begin{lemma}
We have $\hat{\mathcal{C}}^{n+1}(\mathcal{E})=\mathcal{C}^{n+1}(\mathcal{E})$.
\end{lemma}

\begin{proof}
Let $C\in\mathcal{C}^{n+1}(\mathcal{E})$ and let $d_{C}\in Der(R,E^{1})$ be given $\langle d_{C}r,x\rangle =\sigma_{C}(x)r$. We can find $D^{1},...,D^{n}\in\mathfrak{X}$ and $e_{1},...,e_{n}\in E^{1}$ such that $d_{C}(r)=D^{i}(r)e_{i}$. Let $T=C+\nabla_{D^{i}}\wedge e_{i}$. Then, $\sigma_{T}=0$, so $T$ satisfies

\begin{equation}
\langle  T(u),w\rangle+\langle  u,T(w)\rangle=0,
\end{equation}
\begin{equation}
\langle  T(x_{1},...,x_{l},u),w\rangle+\langle  u,T(x_{1},...,x_{l},w)\rangle=0,
\end{equation}
\begin{equation}
\langle  T(x_{1},...x_{i},x_{i+1},...,x_{k})-(-1)^{(|x_{i}|-n)(|x_{i+1}|-n)}T(x_{1},....,x_{i+1},x_{i},...,x_{k}),u\rangle=0.
\end{equation}
In this case, we can define
\begin{equation}
T'\in\oplus_{1\leq j\leq n-1}\oplus_{1\leq k\leq n+1-j}\oplus_{\sum^{k}_{t=1}i_{t}=n+1-j}\mathrm{Hom}_{K}(E^{n-i_{1}}\otimes\cdots\otimes E^{n-i_{k-1}},E^{i_{k}}\wedge E^{j})
\end{equation}
by $T(x_{1},...,x_{k})=:\langle T'(x_{1},...,x_{k-1}),x_{k}\rangle(=\{T'(x_{1},...,x_{k-1}),x_{k}\}_{R})$. Repeating this process, we can get $\tilde{T}\in\mathrm{Sym}(\oplus_{1\leq i\leq n-1}E^{i}[-i])$ of degree $n+1$ such that $T=\mathcal{J}(\tilde{T})$.
\end{proof}

Let  $m\in\mathcal{C}^{n+1}(\mathcal{E})$ with $[m,m]=0$. Then $\delta_{m}=[m,-]$ squares to 0, and we get a subcomplex $\hat{\mathcal{C}}^{\bullet}(\mathcal{E})$. This complex is isomorphic to  $\mathcal{R}^{\bullet}(\mathcal{E})$ with the differential $\delta_{\mathcal{J}^{-1}(m)}=\{\mathcal{J}^{-1}(m),-\}_{R}$.  

When $R=C^{\infty}(M)$ and $E^{i}=\Gamma(M,F^{i})$ for a graded vector bundle $F^{i}\rightarrow M$, this Poisson algebra is isomorphic to the associated dg symplectic manifold$(\mathcal{M},\omega,\Theta)$.

\begin{lemma}
Let $(F^{i}(1\leq i\leq n-1))$ be a graded bundle over a smooth manifold $M$, and $\langle ,\rangle:F^{i}\otimes F^{n-i}\rightarrow C^{\infty}(M)$ a fiberwise non-degenerate graded-symmetric bilinear form. Then degree $n$ graded symplectic manifolds are in one-to-one correspondence with graded vector bundles with $\langle ,\rangle$.
\end{lemma}

\begin{proof}
Any graded manifold is noncanonically diffeomorphic to a graded manifold associated to a graded vector bundle(\cite{bonavolonta2013category},Theorem 1). Let $(\mathcal{M},\omega)$ be a degree $n$ symplectic manifold and let $F^{i}$ the associated graded vector bundle. Then, $E^{n}=\Gamma(TM)$ and the Poisson bracket of degree $-n$ induced by $\omega$ is an extension of $\langle ,\rangle$ as a derivation. (In this case $C^{\infty}(\mathcal{M})\simeq\mathcal{R}^{\bullet}(\mathcal{E})$.)
\end{proof}

\begin{remark}
The diffeomorphism between a graded manifold and a graded manifold associated to a graded vector bundle is noncanonical. Denote the algebra of degree $i$ functions of a graded manifold $\mathcal{M}$ by $\mathcal{A}^{i}$. There exists a short exact sequence
\begin{equation}
\xymatrix{
  0 \ar[r] & (\mathcal{A}^{1})^{2} \ar[r] & \mathcal{A}^{2} \ar[r] & \Gamma(F^{2}) \ar[r] & 0,
}
\end{equation}
where $F^{2}$ is a vector bundle over the base manifold $M$ of $\mathcal{M}$. Fixing a splitting, we can identify $\mathcal{A}^{2}$ with $(\mathcal{A}^{1})^{2}\oplus\Gamma(F^{2})$. For $\mathcal{A}^{i}(i\geq 2)$, we can choose such a splitting.  Thus graded manifolds with a choice of splittings are in one-to-one correspondence with graded vector bundles.

\end{remark}

\begin{theorem}

Let $(R,E^{i}(1\leq i\leq n-1),\langle ,\rangle,d,[-,-])$ be a higher Courant-Dorfman algebra. Suppose $R=C^{\infty}(M)$ for a smooth manifold $M$, and each $E^{i}=\Gamma(F^{i})$ for a graded vector bundle $F^{i}$ over M. Degree $n$ dg symplectic manifolds are in one-to-one correspondence with higher Courant-Dorfman algebras of these types.

\end{theorem}

\begin{proof}
Let $(\mathcal{M},\omega)$ be a degree $n$ symplectic manifold corresponding to $(E^{i},\langle ,\rangle)$, with $\mathcal{A}$ its graded Poisson algebra of polynomial functions. Then $\mathcal{A}^{0}=C^{\infty}(M)$ and $\mathcal{A}^{i}=\mathcal{E}^{i}$ for $1\leq i\leq n-1$, and $\{-,-\}$ restricted to $\mathcal{A}^{i}$ is an extension of $\langle ,\rangle$. Let $\Theta\in\mathcal{A}^{n+1}$ satisfy $\{\Theta,\Theta\}=0$. Given arbitrary $e,e_{1},e_{2}\in\mathcal{A}^{i}$, define a differential $d$ and bracket $[,]$ by
\begin{equation}
d(e)=\{\Theta,e\}, [e_{1},e_{2}]=\{\{e_{1},\Theta\},e_{2}\}.
\end{equation}  
This construction gives a higher Courant-Dorfman algebra.

Conversely, given a higher Courant-Dorfman algebraic structure on $(E^{i},\{,\})$, we can define $\Theta:=\mathcal{J}(\omega_{\phi})$, where $\omega_{\phi}$ is defined as (\ref{diff}) and $\phi$ denotes the bracket of the higher Courant-Dorfman algebras. Locally, $\Theta$ can be written as follows. In a Darboux chart $(\xi^{a(k)})=(q^{a(l)},p^{a(n-l)})(1\leq k\leq n, 1\leq l\leq\lfloor\frac{n}{2}\rfloor)$, corresponding to a chart $(x_{i})$ on $M$ and a local basis $e^{a(k)}$ of sections of $E^{k}$ such that $\langle  e^{a(k)},e^{b(n-k)}\rangle=\delta^{ab}$  
\begin{equation}
\Theta=\sum_{\sum i_{t}=n+1}\phi(q)\xi^{a_{m}(i_{m})}\cdots\xi^{a_{1}{(i_{1})}}
\end{equation}
\begin{equation}
\phi(q)=\langle \cdots\langle [e^{a_{1}(n-i_{1})}, e^{a_{2}(n-i_{2})}],e^{a_{3}(n-i_{3})}\rangle ,\cdots,e^{a_{m}(n-i_{m})}\rangle .
\end{equation}

We can check that $\Theta=\mathcal{J}(\omega_{\phi})$ as follows.
\begin{align*}
\{\Theta,e^{a_{1}(n-i_{1})}\},e^{a_{2}(n-i_{2})}\},\cdots,e^{a_{m}(n-i_{m})}\}&=\langle \cdots\langle [e^{a_{1}(n-i_{1})}, e^{a_{2}(n-i_{2})}],e^{a_{3}(n-i_{3})}\rangle ,\cdots,e^{a_{m}(n-i_{m})}\rangle\\
&=\omega_{\phi}(e^{a_{1}(n-i_{1})}, e^{a_{2}(n-i_{2})},\cdots,e^{a_{m}(n-i_{m})}).
\end{align*}

This satisfies $\{\Theta,\Theta\}=0$ because $\{\Theta,\Theta\}=\{\mathcal{J}(\omega_{\phi}),\mathcal{J}(\omega_{\phi})\}=\mathcal{J}([\omega_{\phi},\omega_{\phi}])=0$.
\end{proof}

\section{Higher PVAs from higher Courant-Dorfman algebras}\label{sec4}

In this section, we define higher Lie conformal algebras and higher Poisson vertex algebras corresponding to higher weak Courant-Dorfman algebras and higher Courant-Dorfman algebras, and check these algebras have a LCA-like or PVA-like property. In particular, a tensor product of a higher PVA and an arbitrary dgca has a structure of degree 0 graded Poisson algebra.

First, we define higher Lie conformal algebras and derive properties of higher weak Courant-Dorfman algebras, in a similar way that we derive the properties of weak Courant-Dorfman algebras from Lie conformal algebras.

\begin{definition}

\textit{A higher Lie conformal algebra} of degree $n$ is a graded $\mathbb{C}[d]$-module $W=\oplus_{m\in\mathbb{Z}_{\geq0}} W^{m}$(i.e. $d$ acts on elements of $W$) with $|d|=1$, which has a degree $1-n$ map which we call $\Lambda$-bracket $[_{\Lambda}]:W\otimes W\rightarrow W[\Lambda]$ with $|\Lambda|=1$ which satisfy the conditions. (Here, $\Lambda$ is an indeterminate.)

\begin{description}

\item[Sesquilinearity]
\begin{equation}
[da_{\Lambda}b]=-\Lambda[a_{\Lambda}b], [a_{\Lambda}db]=-(-1)^{|a|-n}(d+\Lambda)[a_{\Lambda}b],
\end{equation}

\item[skew-symmetry]
\begin{equation}
[a_{\Lambda}b]=-(-1)^{(|a|+1-n)(|b|+1-n)}[b_{-\Lambda-d}a]
\end{equation}
where $d$ acts on the left for $[b_{-\Lambda-d}a]$.

\item[Jacobi identity]
\begin{equation}
[a_{\Lambda}[b_{\Gamma}c]]=[[a_{\Lambda}b]_{\Lambda+\Gamma}c]+(-1)^{(|a|+1-n)(|b|+1-n)}[b_{\Gamma}[a_{\Lambda}c]].
\end{equation}

\end{description}

\end{definition}

We derive the properties of a higher weak Courant-Dorfman algebra from a higher Lie conformal algebra.

The $\Lambda$-bracket is of the form 
\begin{equation}
[a_{\Lambda}b]=\sum_{j\geq0}\Lambda^{j}a_{(j)}b\ (a_{(j)}b\in W^{|a|+|b|+1-n-j}).
\end{equation}
Let
\begin{equation}
[a,b]:=a_{(0)}b,\ \langle a_{\Lambda}b\rangle:=\sum_{j\geq1}\Lambda^{j-1}a_{(j)}b.
\end{equation}
We can write
\begin{equation}
[a_{\Lambda}b]=[a,b]+\Lambda\langle a_{\Lambda}b\rangle.
\end{equation}

Then we derive the properties of a higher Courant-Dorfman algebra by comparing the independent terms of $\Lambda$ on the both sides of the axioms. 

We denote the dependent term of $\Lambda$ by $o(\Lambda)$. From the sesquilinearity, we can get
\begin{equation}
[da,b]+o(\Lambda)=[da_{\Lambda}b]=-\Lambda[a_{\Lambda}b]\Rightarrow[da,b]=0,
\end{equation}
from the skew-symmetry, we can get 
\begin{align}
[a,b]+o(\Lambda)&=[a_{\Lambda}b]=-(-1)^{(|a|+1-n)(|b|+1-n)}[b_{-\Lambda-d}a] \notag \\
&=-(-1)^{(|a|+1-n)(|b|+1-n)}([b,a]-d\langle  b_{-d}a\rangle)+o(\Lambda), \notag \\
(-1)^{(|a|+1-n)(|b|+1-n)}[b,a]+o(\Lambda)&=(-1)^{(|a|+1-n)(|b|+1-n)}[b_{\Lambda}a]=-[a_{-\Lambda-d}b] \notag \\
&=-([a,b]-d\langle  a_{-d}b\rangle)+o(\Lambda) \notag \\
\Rightarrow[a,b]+(-1)^{(|a|+1-n)(|b|+1-n)}[b,a]&=d\langle  a,b\rangle,
\end{align}
where 
\begin{equation}
\langle a,b\rangle:=\frac{1}{2}(\langle  a_{-d}b\rangle+(-1)^{(|a|+1-n)(|b|+1-n)}\langle  b_{-d}a\rangle),
\end{equation}
and from the Jacobi identity, we can get
\begin{align}
[a,[b,c]]+o(\Lambda)+o(\Gamma)&=[a_{\Lambda}[b_{\Gamma}c]]=[[a_{\Lambda}b]_{\Lambda+\Gamma}c]+(-1)^{(|a|+1-n)(|b|+1-n)}[b_{\Gamma}[a_{\Lambda}c]] \notag \\
&=[[a,b],c]+(-1)^{(|a|+1-n)(|b|+1-n)}[b,[a,c]]+o(\Lambda)+o(\Gamma) \notag \\
\Rightarrow&[a,[b,c]]=[[a,b],c]+(-1)^{(|a|+1-n)(|b|+1-n)}[b,[a,c]].
\end{align}

These are properties of a higher weak Courant-Dorfman algebras.

\begin{definition}

\textit{A higher weak Courant-Dorfman algebra} of degree $n$ consists of the following data:

\begin{itemize}

\item a graded vector space $E=(E^{i})$,

\item a graded symmetric bilinear form of degree $-n$ $\langle ,\rangle:E\otimes E\rightarrow E$,

\item a map of degree 1 $d:E\rightarrow E$,

\item a Dorfman bracket of degree $1-n$ $[,]:E\otimes E\rightarrow E$,

\end{itemize}

which satisfies the following conditions.
\begin{equation}
[e_{1},[e_{2},e_{3}]]=[[e_{1},e_{2}],e_{3}]+(-1)^{(|e_{1}|+1-n)(|e_{2}|+1-n)}[e_{2},[e_{1},e_{3}]],
\end{equation}
\begin{equation}
[e_{1},e_{2}]+(-1)^{(|e_{1}|+1-n)(|e_{2}|+1-n)}[e_{2},e_{1}]=d\langle  e_{1},e_{2}\rangle,
\end{equation}
\begin{equation}
[de_{1},e_{2}]=0.
\end{equation}
\end{definition}

Next, we define higher Poisson vertex algebras. We did not assume that $d$ is a differential so far. From now on, we assume $d^{2}=0$. Then, $C=(C^{k},d)$ is a cochain complex. 

\begin{definition}

Let $C=(C^{k},d)$ a cochain complex of vector spaces over $\mathbb{C}$. We say that $C$ is a higher dg Lie conformal algebra of degree $n$ if it endows with a $\Lambda$-bracket $[_{\Lambda}]:C\otimes C\rightarrow C[\Lambda]$, where $|\Lambda|=1$ and $\Lambda^2=0$, defined by
\begin{equation}
a\otimes b\mapsto [a_{\Lambda}b]=a_{(0)}b+\Lambda a_{(1)}b
\end{equation}
satisfying the axioms of higher Lie conformal algebras. $C$ is a higher Poisson vertex algebra of degree $n$ if it is a higher dg LCA and a dgca which satisfies 
\begin{description}
\item[the Leibniz rule]
\begin{equation}
[a_{\Lambda}bc]=[a_{\Lambda}b]c+(-1)^{(|a|+1-n)|b|}b[a_{\Lambda}c].
\end{equation}
\end{description}
\end{definition}

From extended higher Courant-Dorfman algebras, we get the following theorem.

\begin{theorem}

The above higher Poisson vertex algebras generated by elements of degree $0\leq i\leq n-1$ are in one-to-one correspondence with the extended higher Courant-Dorfman algebras whose ground ring is $\mathbb{C}$.

\end{theorem}

\begin{proof}
Assume we have a higher PVA $(C=(C^{k},d),\{_{\Lambda}\})$. We denote $R=C^{0},\mathcal{E}^{i}=C^{i}(1\leq i\leq n-1)$.  $C=(C^{k},d)$ is a dgca, so $R$ is a commutative algebra and each $\mathcal{E}^{i}$ is an $R$-module. We denote the $\Lambda$-bracket by
\begin{equation}\label{116}
a_{(0)}b=[a,b],\ a_{(1)}b=(-1)^{|a|+1-n}\langle a,b\rangle.
\end{equation}

Sesquilinearity says that
\begin{equation}
(da)_{(0)}b+\Lambda(da)_{(1)}b=-\Lambda(a_{(0)}b+\Lambda a_{(1)}b).
\end{equation}

Comparing the 0th-order terms and the first-order terms of $\Lambda$, we have
\begin{equation}
[da,b]=0,\ \langle  da,b\rangle=-(-1)^{|a|-n}[a,b].
\end{equation}

In a similar way, from the skew-symmetry
\begin{equation}
a_{(0)}b+\Lambda a_{(1)}b=-(-1)^{(|a|+1-n)(|b|+1-n)}(b_{(0)}a-(\Lambda+d)b_{(1)}a),
\end{equation}
we can get 
\begin{equation}
\langle  a,b\rangle=-(-1)^{(|a|-n)(|b|-n)}\langle  b,a\rangle,
\end{equation}
\begin{align}
[a,b]+(-1)^{(|a|+1-n)(|b|+1-n)}[b,a]&=(-1)^{(|a|+1-n)(|b|+1-n)+|b|+1-n}d\langle  b,a\rangle \notag \\
&=(-1)^{|a|+1-n}d\langle a,b\rangle.
\end{align}

From the Jacobi identity
\begin{align}\label{hj}
\ &a_{(0)}(b_{(0)}c)+a_{(0)}(\Gamma b_{(1)}c)+\Lambda a_{(1)}(b_{(0)}c)+\Lambda a_{(1)}(\Gamma b_{(1)}c) \notag \\
&=(a_{(0)}b)_{(0)}c+(\Lambda+\Gamma)(a_{(0)}b)_{(1)}c+(\Lambda a_{(1)}b)_{(0)}c+(\Lambda+\Gamma) (\Lambda a_{(1)} b)_{(1)}c \notag \\
&+(-1)^{(|a|+1-n)(|b|+1-n)}\{b_{(0)}(a_{(0)}c)+b_{(0)}(\Lambda a_{(1)}c)+\Gamma b_{(1)}(a_{(0)}c)+\Gamma b_{(1)}(\Lambda a_{(1)}c)\},
\end{align}
we can get
\begin{equation}
[a,[b,c]]=[[a,b],c]+(-1)^{(|a|+1-n)(|b|+1-n)}[b,[a,c]],
\end{equation}
\begin{equation}\label{124}
[a,\langle  b,c\rangle]=\langle [a,b],c\rangle+(-1)^{(|a|+1-n)(|b|-n)}\langle  b,[a,c]\rangle,
\end{equation}
\begin{equation}
\langle  a,\langle  b,c\rangle\rangle=\langle \langle  a,b\rangle,c\rangle+(-1)^{(|a|-n)(|b|-n)}\langle  b,\langle  a,c\rangle\rangle.
\end{equation}

Here we write the detail of the derivation of (\ref{124}) from (\ref{hj}) as an example of how to derive such equalities. Comparing the $\Gamma$ terms in (\ref{hj}), we have
\[
 a_{(0)}(\Gamma b_{(1)}c)=\Gamma (a_{(0)}b)_{(1)}c+(-1)^{(|a|+1-n)(|b|+1-n)}\Gamma b_{(1)}( a_{(0)}c).
\]
Using (\ref{116}), we have
\begin{align*}
[ a,(\Gamma(-1)^{|b|+1-n} \langle b,c\rangle)]&=\Gamma(-1)^{|a|+|b|+1-n+1-n}\langle[a,b],c\rangle \\
&+(-1)^{(|a|+1-n)(|b|+1-n)}\Gamma(-1)^{|b|+1-n}\langle b,[a,c]\rangle.
\end{align*}
Make $\Gamma$ be the left of all terms.
\begin{align*}
(-1)^{|a|+1-n}\Gamma[a,(-1)^{|b|+1-n} \langle b,c\rangle]&=(-1)^{|a|+|b|+1-n+1-n}\Gamma\langle[a,b],c\rangle \\
&+(-1)^{(|a|+1-n)(|b|+1-n)}(-1)^{|b|+1-n}\Gamma\langle b,[a,c]\rangle.
\end{align*}
Devide the both sides by $(-1)^{|a|+1-n}(-1)^{|b|+1-n}=(-1)^{|a|+|b|}$.
\begin{align*}
\Gamma[a,\langle b,c\rangle]&=\Gamma\langle[a,b],c\rangle \\
&+(-1)^{(|a|-n)(|b|+1-n)+|a|+|b|}\Gamma\langle b,[a,c]\rangle.
\end{align*}
From the computation
\begin{align*}
&(-1)^{(|a|-n)(|b|+1-n)+|a|+|b|}=(-1)^{|a||b|+|a|-|a|n-|b|n-n+n^2+|a|+|b|}\\
&=(-1)^{|a||b|+|b|-|b|n-|a|n-n+n^2}=(-1)^{(|a|+1-n)(|b|-n)},
\end{align*}
we have
\[
[a,\langle b,c\rangle]=\langle[a,b],c\rangle+(-1)^{(|a|+1-n)(|b|-n)}\langle b,[a,c]\rangle,
\]
which coincides with (\ref{124}). We can derive other equalities in a similar way.

From the Leibniz rule
\begin{equation}
a_{(0)}(bc)+\Lambda a_{(1)}(bc)=(a_{(0)}b)c+\Lambda (a_{(1)}b)c+(-1)^{(|a|+1-n)|b|}b(a_{(0)}c+\Lambda a_{(1)}c),
\end{equation}
we can get
\begin{equation}
[a,b\cdot c]=[a,b]\cdot c+(-1)^{(|a|+1-n)|b|}b\cdot[a,c],
\end{equation}
\begin{equation}
\langle  a,b\cdot c\rangle=\langle  a,b\rangle\cdot c+(-1)^{(|a|-n)|b|}b\cdot\langle  a,c\rangle.
\end{equation}
The conditions coincide with the definition of extended higher Courant-Dorfman algebras.

Conversely, assuming that we have an extended Courant-Dorfman algebra $\mathcal{E}=(\mathcal{E}^{k},d),\langle ,\rangle,[,]$, define a $\Lambda$-bracket $\{a_{\Lambda}b\}=[a,b]+(-1)^{|a|+1-n}\Lambda\langle  a,b\rangle$. Then, this bracket satisfies the conditions of a $\Lambda$-bracket. Note that we can show the Jacobi identity of a higher Poisson vertex algebra(\ref{hj}) using the sesquilinearity(\ref{ses}), skew-symmetry(\ref{ss}) and the Jacobi identity(\ref{cj}) of a higher Courant-Dorfman algebra.

\end{proof}

Next, we check this algebra has a PVA-like property. In particular, we show we can construct a graded Lie algebra from a tensor product of a higher LCA and an arbitrary dgca and a graded Poisson algebra out of that of a higher PVA and an arbitrary dgca.

\begin{lemma}

Let $C=(C^{k},d_{1})$ be a higher dg LCA and $(E,d_{2})$ be a dgca. Then, the tensor product $C\otimes E$ of cochain complexes is also a higher dg LCA by defining a bracket as
\begin{align*}
[a\otimes f_{\Lambda}b\otimes g]&=(-1)^{(|b|+1-n)|f|}[a_{\Lambda+d_{2f}}b]\otimes fg\\
&=(-1)^{(|b|+1-n)|f|}\{a_{(0)}b\otimes fg+(\Lambda+d_{2f})a_{(1)}b\otimes fg\}\\
&=(-1)^{(|b|+1-n)|f|}\{a_{(0)}b\otimes fg+\Lambda a_{(1)}b\otimes fg+(-1)^{|a|+|b|-n}a_{(1)}b\otimes (d_{2f}f)g\},
\end{align*}
\[
d(a\otimes f)=d_{1}a\otimes f+(-1)^{|a|}a\otimes d_{2f}f
\]
where $d_{2f}$ is the differential on $E$ which acts on $f$.

\end{lemma}

\begin{proof}

sesquilinearity:
\begin{align}
[d(a\otimes f)_{\Lambda}b\otimes g]&=(-1)^{(|b|+1-n)|f|}\{[d_{1}a\otimes f_{\Lambda}b\otimes g]+(-1)^{|a|}[a\otimes d_{2f}f_{\Lambda}b\otimes g]\} \notag \\
&=(-1)^{(|b|+1-n)|f|}\{[d_{1}a_{\Lambda+d_{2f}}b]\otimes fg+(-1)^{|a|+|b|+1-n}[a_{\Lambda+d_{2f}}b]\otimes (d_{2f}f)g\} \notag \\
&=(-1)^{(|b|+1-n)|f|}\{-(\Lambda+d_{2f})[a_{\Lambda+d_{2f}}b]\otimes fg+(-1)^{|a|+|b|+1-n}[a_{\Lambda+d_{2f}}b]\otimes (d_{2f}f)g\} \notag \\
&=(-1)^{(|b|+1-n)|f|}\{-\Lambda[a_{\Lambda+d_{2f}}b]\otimes fg-(-1)^{|a|+|b|+1-n}[a_{\Lambda+d_{2f}}b]\otimes (d_{2f}f)g \notag \\
&+(-1)^{|a|+|b|+1-n}[a_{\Lambda+d_{2f}}b]\otimes (d_{2f}f)g\} \notag \\
&=-\Lambda[a\otimes f_{\Lambda}b\otimes g].
\end{align}

skew-symmetry:

\begin{align}
[a\otimes f_{\Lambda}b\otimes g]&=(-1)^{(|b|+1-n)|f|}[a_{\Lambda+d_{2f}}b]\otimes fg \notag \\
&=-(-1)^{(|a|+1-n)(|b|+1-n)+(|b|+1-n)|f|}[b_{-\Lambda-d_{2f}-d_{1}}a]\otimes fg \notag \\
&=-(-1)^{(|a|+|f|+1-n)(|b|+|g|+1-n)-(|a|+1-n)|g|}[b_{-\Lambda-d_{2f}-d_{1}-d_{2g}+d_{2g}}a]\otimes gf \notag \\
&=-(-1)^{(|a|+|f|+1-n)(|b|+|g|+1-n)-(|a|+1-n)|g|}[b_{-\Lambda-(d_{1}+d_{2f}+d_{2g})+d_{2g}}a]\otimes gf \notag \\
&=-(-1)^{(|a|+|f|+1-n)(|b|+|g|+1-n)-(|a|+1-n)|g|}[b_{-\Lambda-d+d_{2g}}a]\otimes gf \notag \\
&=-(-1)^{(|a|+|f|+1-n)(|b|+|g|+1-n)}[b\otimes g_{-\Lambda-d}a\otimes f]
\end{align}
where $d_{2g}$ is the differential on $E$ which acts on $g$.

Using the Jacobi identity of the original higher LCAs, we can be check the Jacobi identity in a similar way.

\end{proof}

\begin{definition}

A graded Lie algebra $\mathcal{C}$ of degree $N\in\mathbb{Z}$ is a graded vector space with a bilinear operation $[,]:\mathcal{C}\otimes\mathcal{C}\rightarrow\mathcal{C}$ of degree $N$ satisfying:

(1)skew-symmetry:$[a,b]=-(-1)^{(|a|+N)(|b|+N)}[b,a]$,

(2)Jacobi identity:$[a,[b,c]]=[[a,b],c]+(-1)^{(|a|+N)(|b|+N)}[b,[a,c]]$.

\end{definition}

\begin{lemma}

Let $C=(C^{k},d)$ be a higher dg Lie conformal algebra of degree $n$. Then $C/\mathrm{Im}d$ is naturally a graded Lie algebra of  degree $(1-n)$ with bracket
\begin{equation}
[a+dC,b+dC]=[a_{\Lambda}b]_{\Lambda=0}+dC
\end{equation}

\end{lemma}

\begin{proof}

The well-definedness follows from the sesquilinearity.

\begin{equation}
[da,b]=-(\Lambda[a_{\Lambda}b])_{\Lambda=0}=0,
\end{equation}
\begin{equation}
[a,db]=-(-1)^{|a|-n}((\Lambda+d)[a_{\Lambda}b])_{\Lambda=0}=d[a_{\Lambda}b]\simeq0.
\end{equation}

The skew-symmetry follows from the skew-symmetry of the complex.
\begin{align}
[a,b]&=[a_{\Lambda}b]_{\Lambda=0} \notag \\
&=-(-1)^{(|a|+1-n)(|b|+1-n)}[b_{-\Lambda-d}a]_{\Lambda=0} \notag \\
&\simeq-(-1)^{(|a|+1-n)(|b|+1-n)}[b_{\Lambda}a]_{\Lambda=0} \notag \\
&=-(-1)^{(|a|+1-n)(|b|+1-n)}[b,a]
\end{align}

In the similar way, we can check the Jacobi identity follows from the Jacobi identity of the complex.

\end{proof}

\begin{lemma}

Let $L$ be a graded Lie algebra of degree $N$. Then, $L[-N]$ is a graded Lie algebra of degree 0 with the same bracket.
\end{lemma}

\begin{proof}

It satisfies the skew-symmetry and the Jacobi identity due to the grade-shifting.

\end{proof}

For any higher dg LCA of degree $n$ $C=(C^{k},d)$ and dgca $E=(E^{k},D)$, we put $L(C,E)=C\otimes E/\mathrm{Im}(d+D)$ and $\mathrm{Lie}(C,E)=L(C,E)[n-1]$. By the above lemmas, $\mathrm{Lie}(C,E)$ is a graded Lie algebra of weight 0 via
\begin{equation}
\{a\otimes f,b\otimes g\}=(-1)^{(|b|+1-n)|f|}(a_{(0)}b\otimes fg+(-1)^{(|a|+|b|-n)}a_{(1)}b\otimes (Df)g).
\end{equation}  

Next, we discuss the Poisson algebraic structure. Let $C=(C^{k},d)$ be a higher PVA of degree $n$ and $E=(E^{k},D)$ be a dgca. Then, $C\otimes E[n-1]$ is a dgca with products $a\otimes f\cdot b\otimes g=(-1)^{|b||f|}a\cdot b\otimes f\cdot g$, and $\mathrm{Lie}(C,E)$ is a graded Lie algebra. Let $I_{d+D}$ denote the ideal generated by $\mathrm{Im}(d+D)$, and we put 
\begin{equation}
P(C,E)=C\otimes E[n-1]/I_{d+D}.
\end{equation}

\begin{theorem}
$P(C,E)$ is a graded Poisson algebra with
\begin{equation}
[a\otimes f]\cdot[b\otimes g]=(-1)^{|b||f|}[a\cdot b\otimes fg],
\end{equation}
\begin{equation}
\{[a\otimes f],[b\otimes g]\}=(-1)^{(|b|+1-n)|f|}(a_{(0)}b\otimes fg+(-1)^{(|a|+|b|-n)}a_{(1)}b\otimes (df)g).
\end{equation}

\end{theorem}

\begin{proof}

If $a,b\in I_{d+D}$, then $a\cdot b,(d+D)a\in I_{d+D}$. Therefore, $I_{d+D}$ is a dg ideal of $C\otimes E[n-1]$ and $P(C,E)$ is a dgca. If $a,b\in I_{d+D}/\mathrm{Im}(d+D)$, then $[a,b]\in I_{d+D}/\mathrm{Im}(d+D)$ by the Leibiniz identity of $C$, so $I_{d+D}/Im (d+D)$ is a graded Lie ideal of $Lie(C,E)$ and $P(C,E)$ is a Lie algebra with the Lie bracket. The Leibiniz identity follows from the Leibniz identity of $C$. So $P(C,E)$ is a Poisson algebra.

\end{proof}

By the above theorem, we get a graded Poisson algebra from a higher PVA and a dgca. Using the notations of a higher Courant-Dorfman algebra, the Poisson bracket becomes

\begin{align*}
\{a\otimes f,b\otimes g\}&=(-1)^{(|b|+1-n)|f|}(a_{(0)}b\otimes fg+(-1)^{(|a|+|b|-n)}a_{(1)}b\otimes (Df)g) \\
&=(-1)^{(|b|+1-n)|f|}([a,b]\otimes fg+(-1)^{(|b|+1)}\langle a,b\rangle\otimes (Df)g).
\end{align*}

\begin{example}

We define the BFV analog of formal distribution Lie algebras. Define the algebra of power series

\begin{equation}
\mathbb{C}[[t_{1},t^{-1}_{1},...t_{n},t^{-1}_{n}]][\theta_{1},...,\theta_{n}]
\end{equation}
where $t_{i}$ are even coordinates of degree 0, $\theta_{i}$ are odd coordinates of degree 1. 

Define the "de-Rham differential" as
\begin{equation}
df:=\sum_{i}\frac{\partial f}{\partial t^{i}}\theta_{i}.
\end{equation}

Let $C=(C^{n},Q)$ be a cochain complex which is a higher dg LCA of degree $n+1$. For 
\begin{equation}
V:=C\otimes\mathbb{C}\left[[t_{1},t^{-1}_{1},...t_{n},t^{-1}_{n}]][\theta_{1},...,\theta_{n}\right][n]/(Q\alpha\otimes f+\alpha\otimes df),
\end{equation}
the bracket

\begin{align}
&[a\otimes t_{1}^{p_{1}}\cdots t_{n}^{p_{n}}\theta^{J},b\otimes t_{1}^{q_{1}}\cdots t_{n}^{q_{n}}\theta^{K}] \notag \\
&=(-1)^{(|b|+1-n)|J|}\{(a_{(0)}b)t_{1}^{p_{1}+q_{1}}\cdots t_{n}^{p_{n}+q_{n}}\theta^{J\cdot K}+(-1)^{|a|+|b|-n}\sum^{n}_{k=1}(a_{(1)}b)p_{k}t_{1}^{p_{1}+q_{1}}\cdots t_{k}^{p_{k}+q_{k}-1} t_{n}^{p_{n}+q_{n}}\theta^{J\cdot\{k\}\cdot K}\},
\end{align}
\begin{equation*}
J,K\subset\{1,...,n\},J\cdot K=\left\{\begin{array}{ll}
\emptyset&(J\cap K\neq\emptyset),\\
J\cup K&(J\cap K=\emptyset),
\end{array}\right.
\end{equation*}
makes the graded Lie algeraic structure.

We define a formal distribution, 
\begin{align}
a(Z_{1},...,Z_{n})&=a(z_{1},...,z_{n},\zeta_{1},...,\zeta_{n}) \notag \\
&=\sum_{m_{i}\in\mathbb{Z},J\subset\{1,...,n\}}z_{1}^{-1-m_{1}}\cdots z_{n}^{-1-m_{n}}\zeta^{\{1,...,n\}\backslash J}\alpha t_{1}^{m_{1}}\cdots t_{n}^{m_{n}}\theta^{J},
\end{align}

and the formal $\delta$-function, 
\begin{align}
\delta(Z-W)&=\delta(z_{1}-w_{1})\cdots\delta({z_{n}-w_{n}})\delta({\zeta_{1}-\xi_{1}})\cdots\delta(\zeta_{n}-\xi_{n}) \notag \\
&=\sum_{m_{i}\in\mathbb{Z}}z_{1}^{-m_{1}-1}w_{1}^{m_{1}}\cdots z_{n}^{-m_{n}-1}w_{n}^{m_{n}}(\zeta_{1}-\xi_{1})\cdots(\zeta_{n}-\xi_{n}),  
\end{align}

Then, we get 
\begin{equation}
[a(Z),b(W)]=[a,b](W)\delta(Z-W)+\langle a,b\rangle (W)d(\delta(Z-W)),
\end{equation}
where $[a,b]:=(-1)^{(|b|+1-n)n}a_{(0)}b$ and $\langle a,b\rangle:=(-1)^{(|b|+1-n)n+|a|+|b|-n}a_{(1)}b$.

(For another example of formal distribution Lie algebras using superfields, see \cite{heluani2006supersymmetric}.)

 Consider the case $n=2$. Let $C=(C^{n},Q)$ be a higher PVA of degree $2$. Then $P(C,\mathbb{C}[[t,t^{-1}]][\theta])$ is a graded Poisson algebra via
\begin{equation}
\{a t^{m},b t^{n}\}=(a_{(0)}b)t^{m+n}+(a_{(1)}b)mt^{m+n-1}\theta,\ \{a t^{m}\theta,b t^{n}\}=(a_{(0)}b)t^{m+n}\theta.
\end{equation} 

An extended higher Courant-Dorfman algebra of degree 2 is the same as a Courant-Dorfman algebra, and there is a PVA corresponding to a given higher PVA. We denote the PVA by $\tilde{C}$. We restrict $P(C,\mathbb{C}[[t,t^{-1}]][\theta])$ to the degree 0 part. Explicitly,
\begin{equation}
P(C,\mathbb{C}[[t,t^{-1}]][\theta])|_{degree 0}=\{a t^{m_{1}},b t^{m_{2}}\theta|a\in C^{1},b\in C^{0},m_{1},m_{2}\in\mathbb{Z}\}.
\end{equation}

$P(C,\mathbb{C}[[t,t^{-1}]][\theta])|_{degree 0}$ is isomorphic to the Poisson algebra arising from the associated Poisson vertex algebra. 
This subalgebra corresponds to the physical current algebra of a BFV current algebra.  
\end{example}

\begin{example}

Let $(\mathcal{M},\omega,Q=\{\Theta,-\})$ be a degree $n$ dg symplectic manifold and $C=C^{n-1}(C^{\infty}(\mathcal{M}))=\{a\in C^{\infty}(\mathcal{M}):|a|\leq n-1\}$ and consider a higher Courant-Dorfman algebra on $C$. Let $\Sigma_{n-1}$ be a $n-1$ dimensional manifold and $E=(\Omega^{\bullet}(\Sigma_{n-1}),D)$ be its de-Rham complex. Then, $P(C,E)$ is equipped with degree 0 Poisson bracket with
\begin{equation}
\label{HP}
[a\otimes\epsilon_{1},b\otimes\epsilon_{2}]=(-1)^{(|b|+1-n)|f|}(\{\{a,\Theta\},b\}\otimes \epsilon_{1}\epsilon_{2}+(-1)^{|b|+1}\{a,b\}\otimes (D\epsilon_{1})\epsilon_{2}),
\end{equation} 
where $a,b\in C$ and $\epsilon_{1},\epsilon_{2}\in E$. This is an algebraic description of BFV current algebras from dg symplectic manifolds\cite{ikeda2013currentdg},\cite{arvanitakis2021brane}.

BFV current algebras are Poisson brackets on $C^{\infty}(Map(T[1]\Sigma_{n-1},\mathcal{M}))$, where $T[1]\Sigma_{n-1}$ is the shifted tangent space of $\Sigma_{n-1}$. In order to get the currents, we have to take a proper Lagrangian submanifold of $Map(T[1]\Sigma_{n-1},\mathcal{M})$. One way is the zero-locus reduction\cite{grigoriev2001becchi}.

\begin{proposition}[{\cite[Proposition 1]{arvanitakis2021brane}}]

We take a degree $-n$ graded Poisson algebra $P$ with a differential $Z$, and take a quotient $P/I_{Z}$, where $I_{Z}$ is the ideal of $P$ generated by $Z$-exact terms.

Then, $P/I_{Z}$ is a degree $-n+1$ Poisson algebra with the derived bracket
\begin{equation}
\{[a],[b]\}=[\{a,Z(b)\}].
\end{equation}  

\end{proposition}

Applying to the BFV current algebras we get the Poisson bracket on 

$C^{\infty}(Map(T[1]\Sigma_{n-1},\mathcal{M}))/I_{\tilde{D}+\tilde{Q}}$, where $\tilde{D}$ and $\tilde{Q}$ is the differential on $Map(T[1]\Sigma_{n-1},\mathcal{M})$ induced by $D$ and $Q$. 

For $a\in C^{\infty}(\mathcal{M})$ and $\epsilon\in C^{\infty}(T[1]\Sigma_{n-1})$, We define $J_{\epsilon}\left(a\right)\in C^{\infty}(Map(T[1]\Sigma_{n-1},\mathcal{M}))$ by
\begin{equation}
J_{\epsilon}\left(a\right)(\phi)=\int_{T[1]\Sigma_{n-1}}\epsilon\cdot\phi^{*}(a)(\sigma,\theta)d^{n-1}\sigma d^{n-1}\theta,
\end{equation} 
where $\epsilon\in C^{\infty}(\Sigma_{n-1})$ are test functions on $T[1]\Sigma_{n-1}$, $\sigma,\theta$ are coordinates on $T[1]\Sigma_{n-1}$ of degree 0 and 1, $\phi\in Map(T[1]\Sigma_{n-1},\mathcal{M})$ and $\phi^{*}(a)$ is the pullback of $a$. Then the Poisson bracket is of the form
\begin{align}
\label{CA}
\{J_{\epsilon_{1}}\left(a\right),J_{\epsilon_{2}}\left(b\right)\}(\phi)&=\int_{T[1]\Sigma_{n-1}}\epsilon_{1}\epsilon_{2}\cdot\phi^{*}(\{\{a,\Theta\},b\})(\sigma,\theta)d^{n-1}\sigma d^{n-1}\theta \notag \\
&+\int_{T[1]\Sigma_{n-1}}(D\epsilon_{1})\epsilon_{2}\cdot\phi^{*}(\{a,b\})(\sigma,\theta)d^{n-1}\sigma d^{n-1}\theta,
\end{align}
where $\epsilon_{1},\epsilon_{2}\in C^{\infty}(T[1]\Sigma_{n-1})$ are test functions on $T[1]\Sigma_{n-1}$, $\sigma,\theta$ are coordinates on $T[1]\Sigma_{n-1}$ of degree 0 and 1,

Comparing to ($\ref{HP}$) and ($\ref{CA}$), we see that taking the quotient corresponds to the zero-locus reduction.

\end{example}

\section{Outlooks}\label{sec5}

In this paper, we gave higher analogs of Lie conformal algebras and Poisson vertex algebras. It is natural to ask whether they have the same applications as ordinary Lie conformal algebras and Poisson vertex algebras. For example, our higher PVAs may be used to analyze multi-variable Hamiltonian PDEs. Considering the algebraic description of more general currents would be important.

Another interesting problem is the non-commutative analog. In \cite{alvarez2023noncommutative}, non-commutative version of Courant-Dorfman algebras and Poisson vertex algebras, which are called double Courant-Dorfman algebras and double Poisson vertex algebras, are considered, and the one-to-one correspondence between double Courant-Dorfman algebras and double Poisson vertex algebras are proved. It may be possible to give the higher analog of this relation (higher double PVAs and higher double Courant-Dorfman algebras) in terms of dg bisymplectic algebras, and it would be meaningful on noncommutative-geometric areas. Another way of taking the non-commutative version is the quantization, in analogy with vertex algebras. In particular, the higher analog of 1-truncated vertex algebras or ``higher vertex algebroid'' would be given by deforming our algebras.        

\bmhead{Acknowledgements}

The author is grateful to Prof. Noriaki Ikeda and Prof. Hidetoshi Awata for helpful comments and fruitful discussions. The author also thanks to Prof. Roberto Rubio for his comments. The author is thankful to the anonymous referee for detailed comments and valuable suggestions.

\section*{Declarations}

The author confirms that the data supporting the findings of this study are available within the article. The author declares that he has no conflict of interest.

\bibliography{sn-bibliography}


\begin{thebibliography}{30}
\ifx \bisbn   \undefined \def \bisbn  #1{ISBN #1}\fi
\ifx \binits  \undefined \def \binits#1{#1}\fi
\ifx \bauthor  \undefined \def \bauthor#1{#1}\fi
\ifx \batitle  \undefined \def \batitle#1{#1}\fi
\ifx \bjtitle  \undefined \def \bjtitle#1{#1}\fi
\ifx \bvolume  \undefined \def \bvolume#1{\textbf{#1}}\fi
\ifx \byear  \undefined \def \byear#1{#1}\fi
\ifx \bissue  \undefined \def \bissue#1{#1}\fi
\ifx \bfpage  \undefined \def \bfpage#1{#1}\fi
\ifx \blpage  \undefined \def \blpage #1{#1}\fi
\ifx \burl  \undefined \def \burl#1{\textsf{#1}}\fi
\ifx \doiurl  \undefined \def \doiurl#1{\url{https://doi.org/#1}}\fi
\ifx \betal  \undefined \def \betal{\textit{et al.}}\fi
\ifx \binstitute  \undefined \def \binstitute#1{#1}\fi
\ifx \binstitutionaled  \undefined \def \binstitutionaled#1{#1}\fi
\ifx \bctitle  \undefined \def \bctitle#1{#1}\fi
\ifx \beditor  \undefined \def \beditor#1{#1}\fi
\ifx \bpublisher  \undefined \def \bpublisher#1{#1}\fi
\ifx \bbtitle  \undefined \def \bbtitle#1{#1}\fi
\ifx \bedition  \undefined \def \bedition#1{#1}\fi
\ifx \bseriesno  \undefined \def \bseriesno#1{#1}\fi
\ifx \blocation  \undefined \def \blocation#1{#1}\fi
\ifx \bsertitle  \undefined \def \bsertitle#1{#1}\fi
\ifx \bsnm \undefined \def \bsnm#1{#1}\fi
\ifx \bsuffix \undefined \def \bsuffix#1{#1}\fi
\ifx \bparticle \undefined \def \bparticle#1{#1}\fi
\ifx \barticle \undefined \def \barticle#1{#1}\fi
\bibcommenthead
\ifx \bconfdate \undefined \def \bconfdate #1{#1}\fi
\ifx \botherref \undefined \def \botherref #1{#1}\fi
\ifx \url \undefined \def \url#1{\textsf{#1}}\fi
\ifx \bchapter \undefined \def \bchapter#1{#1}\fi
\ifx \bbook \undefined \def \bbook#1{#1}\fi
\ifx \bcomment \undefined \def \bcomment#1{#1}\fi
\ifx \oauthor \undefined \def \oauthor#1{#1}\fi
\ifx \citeauthoryear \undefined \def \citeauthoryear#1{#1}\fi
\ifx \endbibitem  \undefined \def \endbibitem {}\fi
\ifx \bconflocation  \undefined \def \bconflocation#1{#1}\fi
\ifx \arxivurl  \undefined \def \arxivurl#1{\textsf{#1}}\fi
\csname PreBibitemsHook\endcsname

\bibitem[\protect\citeauthoryear{Courant}{1990}]{C90}
\begin{barticle}
\bauthor{\bsnm{Courant}, \binits{T.}}:
\batitle{Dirac manifolds}.
\bjtitle{Trans. Am. Math. Soc}
\bvolume{319},
\bfpage{631}--\blpage{661}
(\byear{1990})
\doiurl{10.1090/S0002-9947-1990-0998124-1}
\end{barticle}
\endbibitem

\bibitem[\protect\citeauthoryear{Liu et~al.}{1997}]{liu1997manin}
\begin{barticle}
\bauthor{\bsnm{Liu}, \binits{Z.-J.}},
\bauthor{\bsnm{Weinstein}, \binits{A.}},
\bauthor{\bsnm{Xu}, \binits{P.}}:
\batitle{Manin triples for {L}ie bialgebroids}.
\bjtitle{Journal of Differential Geometry}
\bvolume{45}(\bissue{3}),
\bfpage{547}--\blpage{574}
(\byear{1997})
\end{barticle}
\endbibitem

\bibitem[\protect\citeauthoryear{VG}{1990}]{vg1990hamiltonian}
\begin{barticle}
\bauthor{\bsnm{Drinfeld}, \binits{V.G,}}:
\batitle{Hamiltonian structures on {L}ie groups, {L}ie bialgebras and the geometric
  meaning of the classical {Y}ang-{B}axter equations}.
\bjtitle{Yang-Baxter Equation in Integrable Systems}
\bvolume{10}(\bissue{2}),
\bfpage{222}
(\byear{1990})
\end{barticle}
\endbibitem

\bibitem[\protect\citeauthoryear{Gualtieri}{2003}]{Gualtieri:2003dx}
\begin{botherref}
\oauthor{\bsnm{Gualtieri}, \binits{M.}}:
{Generalized complex geometry}.
PhD thesis,
Oxford U.
(2003)
\end{botherref}
\endbibitem

\bibitem[\protect\citeauthoryear{Cavalcanti and
  Gualtieri}{2011}]{cavalcanti2011generalized}
\begin{botherref}
\oauthor{\bsnm{Cavalcanti}, \binits{G.R.}},
\oauthor{\bsnm{Gualtieri}, \binits{M.}}:
Generalized complex geometry and {T}-duality.
arXiv preprint arXiv:1106.1747
(2011)
\end{botherref}
\endbibitem

\bibitem[\protect\citeauthoryear{Roytenberg}{2007}]{roytenberg2007aksz}
\begin{barticle}
\bauthor{\bsnm{Roytenberg}, \binits{D.}}:
\batitle{A{K}{S}{Z}--{B}{V} formalism and {C}ourant algebroid-induced topological field
  theories}.
\bjtitle{Letters in Mathematical Physics}
\bvolume{79},
\bfpage{143}--\blpage{159}
(\byear{2007})
\end{barticle}
\endbibitem

\bibitem[\protect\citeauthoryear{Coimbra
  et~al.}{2011}]{coimbra2011supergravity}
\begin{barticle}
\bauthor{\bsnm{Coimbra}, \binits{A.}},
\bauthor{\bsnm{Strickland-Constable}, \binits{C.}},
\bauthor{\bsnm{Waldram}, \binits{D.}}:
\batitle{Supergravity as generalised geometry {I}: type {I}{I} theories}.
\bjtitle{Journal of High Energy Physics}
\bvolume{2011}(\bissue{11}),
\bfpage{1}--\blpage{35}
(\byear{2011})
\end{barticle}
\endbibitem

\bibitem[\protect\citeauthoryear{Vaisman}{2013}]{vaisman2013towards}
\begin{botherref}
\oauthor{\bsnm{Vaisman}, \binits{I.}}:
Towards a double field theory on para-{H}ermitian manifolds.
Journal of Mathematical Physics
\textbf{54}(12)
(2013)
\end{botherref}
\endbibitem

\bibitem[\protect\citeauthoryear{Roytenberg}{2002}]{roytenberg2002structure}
\begin{barticle}
\bauthor{\bsnm{Roytenberg}, \binits{D.}}:
\batitle{On the structure of graded symplectic supermanifolds and {C}ourant
  algebroids}.
\bjtitle{Contemporary Mathematics}
\bvolume{315},
\bfpage{169}--\blpage{186}
(\byear{2002})
\end{barticle}
\endbibitem

\bibitem[\protect\citeauthoryear{Roytenberg}{2009}]{roytenberg2009courant}
\begin{barticle}
\bauthor{\bsnm{Roytenberg}, \binits{D.}}:
\batitle{Courant--{D}orfman algebras and their cohomology}.
\bjtitle{Letters in Mathematical Physics}
\bvolume{90}(\bissue{1}),
\bfpage{311}--\blpage{351}
(\byear{2009})
\end{barticle}
\endbibitem

\bibitem[\protect\citeauthoryear{Knizhnik and
  Zamolodchikov}{1984}]{knizhnik1984current}
\begin{barticle}
\bauthor{\bsnm{Knizhnik}, \binits{V.G.}},
\bauthor{\bsnm{Zamolodchikov}, \binits{A.B.}}:
\batitle{Current algebra and {W}ess-{Z}umino model in two dimensions}.
\bjtitle{Nuclear Physics B}
\bvolume{247}(\bissue{1}),
\bfpage{83}--\blpage{103}
(\byear{1984})
\end{barticle}
\endbibitem

\bibitem[\protect\citeauthoryear{Alekseev and
  Strobl}{2005}]{alekseev2005current}
\begin{barticle}
\bauthor{\bsnm{Alekseev}, \binits{A.}},
\bauthor{\bsnm{Strobl}, \binits{T.}}:
\batitle{Current algebras and differential geometry}.
\bjtitle{Journal of High Energy Physics}
\bvolume{2005}(\bissue{03}),
\bfpage{035}
(\byear{2005})
\end{barticle}
\endbibitem

\bibitem[\protect\citeauthoryear{Ekstrand and
  Zabzine}{2011}]{ekstrand2011courant}
\begin{barticle}
\bauthor{\bsnm{Ekstrand}, \binits{J.}},
\bauthor{\bsnm{Zabzine}, \binits{M.}}:
\batitle{Courant-like brackets and loop spaces}.
\bjtitle{Journal of High Energy Physics}
\bvolume{2011}(\bissue{3}),
\bfpage{1}--\blpage{17}
(\byear{2011})
\end{barticle}
\endbibitem

\bibitem[\protect\citeauthoryear{Ekstrand}{2011}]{ekstrand2011going}
\begin{botherref}
\oauthor{\bsnm{Ekstrand}, \binits{J.}}:
Going round in circles: from sigma models to vertex algebras and back.
PhD thesis,
Acta Universitatis Upsaliensis
(2011)
\end{botherref}
\endbibitem

\bibitem[\protect\citeauthoryear{Barakat et~al.}{2009}]{barakat2009poisson}
\begin{barticle}
\bauthor{\bsnm{Barakat}, \binits{A.}},
\bauthor{\bsnm{De~Sole}, \binits{A.}},
\bauthor{\bsnm{Kac}, \binits{V.G.}}:
\batitle{Poisson vertex algebras in the theory of {H}amiltonian equations}.
\bjtitle{Japanese Journal of Mathematics}
\bvolume{4}(\bissue{2}),
\bfpage{141}--\blpage{252}
(\byear{2009})
\end{barticle}
\endbibitem

\bibitem[\protect\citeauthoryear{Ikeda and Koizumi}{2013}]{ikeda2013current}
\begin{barticle}
\bauthor{\bsnm{Ikeda}, \binits{N.}},
\bauthor{\bsnm{Koizumi}, \binits{K.}}:
\batitle{Current algebras and {Q}{P}-manifolds}.
\bjtitle{International Journal of Geometric Methods in Modern Physics}
\bvolume{10}(\bissue{06}),
\bfpage{1350024}
(\byear{2013})
\end{barticle}
\endbibitem

\bibitem[\protect\citeauthoryear{Bonelli and
  Zabzine}{2005}]{bonelli2005current}
\begin{barticle}
\bauthor{\bsnm{Bonelli}, \binits{G.}},
\bauthor{\bsnm{Zabzine}, \binits{M.}}:
\batitle{From current algebras for p-branes to topological {M}-theory}.
\bjtitle{Journal of High Energy Physics}
\bvolume{2005}(\bissue{09}),
\bfpage{015}
(\byear{2005})
\end{barticle}
\endbibitem

\bibitem[\protect\citeauthoryear{Hatsuda and
  Kimura}{2012}]{hatsuda2012canonical}
\begin{barticle}
\bauthor{\bsnm{Hatsuda}, \binits{M.}},
\bauthor{\bsnm{Kimura}, \binits{T.}}:
\batitle{Canonical approach to {C}ourant brackets for d-branes}.
\bjtitle{Journal of High Energy Physics}
\bvolume{2012}(\bissue{6}),
\bfpage{1}--\blpage{26}
(\byear{2012})
\end{barticle}
\endbibitem

\bibitem[\protect\citeauthoryear{Ikeda and Xu}{2013}]{ikeda2013currentdg}
\begin{botherref}
\oauthor{\bsnm{Ikeda}, \binits{N.}},
\oauthor{\bsnm{Xu}, \binits{X.}}:
Current algebras from dg symplectic pairs in supergeometry.
arXiv preprint arXiv:1308.0100
(2013)
\end{botherref}
\endbibitem

\bibitem[\protect\citeauthoryear{Bessho et~al.}{2016}]{bessho2016topological}
\begin{barticle}
\bauthor{\bsnm{Bessho}, \binits{T.}},
\bauthor{\bsnm{Heller}, \binits{M.A.}},
\bauthor{\bsnm{Ikeda}, \binits{N.}},
\bauthor{\bsnm{Watamura}, \binits{S.}}:
\batitle{Topological membranes, current algebras and {H}-flux-{R}-flux duality
  based on {C}ourant algebroids}.
\bjtitle{Journal of High Energy Physics}
\bvolume{2016}(\bissue{4}),
\bfpage{1}--\blpage{41}
(\byear{2016})
\end{barticle}
\endbibitem

\bibitem[\protect\citeauthoryear{Arvanitakis}{2021}]{arvanitakis2021brane}
\begin{barticle}
\bauthor{\bsnm{Arvanitakis}, \binits{A.S.}}:
\batitle{Brane current algebras and generalised geometry from {Q}{P} manifolds.
  or,“when they go high, we go low”}.
\bjtitle{Journal of High Energy Physics}
\bvolume{2021}(\bissue{11}),
\bfpage{1}--\blpage{40}
(\byear{2021})
\end{barticle}
\endbibitem

\bibitem[\protect\citeauthoryear{Batalin and
  Vilkovisky}{1977}]{batalin1977relativistic}
\begin{barticle}
\bauthor{\bsnm{Batalin}, \binits{I.A.}},
\bauthor{\bsnm{Vilkovisky}, \binits{G.A.}}:
\batitle{Relativistic {S}-matrix of dynamical systems with boson and fermion
  constraints}.
\bjtitle{Physics Letters B}
\bvolume{69}(\bissue{3}),
\bfpage{309}--\blpage{312}
(\byear{1977})
\end{barticle}
\endbibitem

\bibitem[\protect\citeauthoryear{Batalin and
  Fradkin}{1983}]{batalin1983generalized}
\begin{barticle}
\bauthor{\bsnm{Batalin}, \binits{I.A.}},
\bauthor{\bsnm{Fradkin}, \binits{E.S.}}:
\batitle{A generalized canonical formalism and quantization of reducible gauge
  theories}.
\bjtitle{Physics Letters B}
\bvolume{122}(\bissue{2}),
\bfpage{157}--\blpage{164}
(\byear{1983})
\end{barticle}
\endbibitem

\bibitem[\protect\citeauthoryear{Keller and
  Waldmann}{2015}]{keller2015deformation}
\begin{barticle}
\bauthor{\bsnm{Keller}, \binits{F.}},
\bauthor{\bsnm{Waldmann}, \binits{S.}}:
\batitle{Deformation theory of {C}ourant algebroids via the {R}othstein algebra}.
\bjtitle{Journal of Pure and Applied Algebra}
\bvolume{219}(\bissue{8}),
\bfpage{3391}--\blpage{3426}
(\byear{2015})
\end{barticle}
\endbibitem

\bibitem[\protect\citeauthoryear{Bursztyn et~al.}{2019}]{bursztyn2019higher}
\begin{barticle}
\bauthor{\bsnm{Bursztyn}, \binits{H.}},
\bauthor{\bsnm{Martinez~Alba}, \binits{N.}},
\bauthor{\bsnm{Rubio}, \binits{R.}}:
\batitle{On higher {D}irac structures}.
\bjtitle{International Mathematics Research Notices}
\bvolume{2019}(\bissue{5}),
\bfpage{1503}--\blpage{1542}
(\byear{2019})
\end{barticle}
\endbibitem

\bibitem[\protect\citeauthoryear{Voronov}{2005}]{voronov2005higher}
\begin{barticle}
\bauthor{\bsnm{Voronov}, \binits{T.}}:
\batitle{Higher derived brackets and homotopy algebras}.
\bjtitle{Journal of pure and applied algebra}
\bvolume{202}(\bissue{1-3}),
\bfpage{133}--\blpage{153}
(\byear{2005})
\end{barticle}
\endbibitem

\bibitem[\protect\citeauthoryear{Bonavolonta and
  Poncin}{2013}]{bonavolonta2013category}
\begin{barticle}
\bauthor{\bsnm{Bonavolonta}, \binits{G.}},
\bauthor{\bsnm{Poncin}, \binits{N.}}:
\batitle{On the category of {L}ie n-algebroids}.
\bjtitle{Journal of geometry and physics}
\bvolume{73},
\bfpage{70}--\blpage{90}
(\byear{2013})
\end{barticle}
\endbibitem

\bibitem[\protect\citeauthoryear{Heluani and
  Kac}{2006}]{heluani2006supersymmetric}
\begin{botherref}
\oauthor{\bsnm{Heluani}, \binits{R.}},
\oauthor{\bsnm{Kac}, \binits{V.G.}}:
Supersymmetric vertex algebras.
arXiv preprint math/0603633
(2006)
\end{botherref}
\endbibitem

\bibitem[\protect\citeauthoryear{Grigoriev et~al.}{2001}]{grigoriev2001becchi}
\begin{barticle}
\bauthor{\bsnm{Grigoriev}, \binits{M.}},
\bauthor{\bsnm{Semikhatov}, \binits{A.}},
\bauthor{\bsnm{Tipunin}, \binits{I.Y.}}:
\batitle{Becchi--{R}ouet--{S}tora--{T}yutin formalism and zero locus reduction}.
\bjtitle{Journal of Mathematical Physics}
\bvolume{42}(\bissue{8}),
\bfpage{3315}--\blpage{3333}
(\byear{2001})
\end{barticle}
\endbibitem

\bibitem[\protect\citeauthoryear{{\'A}lvarez-C{\'o}nsul
  et~al.}{2023}]{alvarez2023noncommutative}
\begin{barticle}
\bauthor{\bsnm{{\'A}lvarez-C{\'o}nsul}, \binits{L.}},
\bauthor{\bsnm{Fern{\'a}ndez}, \binits{D.}},
\bauthor{\bsnm{Heluani}, \binits{R.}}:
\batitle{Noncommutative {P}oisson vertex algebras and {C}ourant--{D}orfman algebras}.
\bjtitle{Advances in Mathematics}
\bvolume{433},
\bfpage{109269}
(\byear{2023})
\end{barticle}
\endbibitem

\end{thebibliography}

\end{document}